# Natural Hyperbolicity of Hexagonal Boron Nitride in the Deep Ultraviolet

Bongjun Choi[1], Jason Lynch[1], Wangleong Chen[2], Seong-Joon Jeon[3], Hyungseob Cho[3], Kyungmin Yang[1], Jonghwan Kim[3,4,5*], Nader Engheta[1,2,6,7*], Deep Jariwala[1,2*]

[1]Department of Electrical and Systems Engineering, University of Pennsylvania, Philadelphia, Pennsylvania 19104, United States

[2]Department of Materials Science and Engineering, University of Pennsylvania, Philadelphia, Pennsylvania 19104, United States

[3]Department of Materials Science and Engineering, Pohang University of Science and Technology, Pohang, Republic of Korea.

[4]Department of Physics, Pohang University of Science and Technology, Pohang, Republic of Korea.

[5]Center for van der Waals Quantum Solids, Institute for Basic Science (IBS), Pohang, Republic of Korea

[6]Department of Bioengineering, University of Pennsylvania, Philadelphia, Pennsylvania 19104, United States

[7]Department of Physics and Astronomy, University of Pennsylvania, Philadelphia, Pennsylvania, 19104, United States

[*] Corresponding authors: dmj@seas.upenn.edu, engheta@seas.upenn.edu, jonghwankim@postech.ac.kr

## Abstract

Hyperbolic media enable unique optical phenomena including hyperlensing, negative refraction, enhanced photonic density of states (PDOS), and highly confined polaritons. While most hyperbolic media are artificially engineered metamaterials, certain natural materials with extreme anisotropy can exhibit hyperbolic dispersion. Here, we report the first observation of natural hyperbolic dispersion in hexagonal boron nitride (hBN) in the deep-ultraviolet (DUV) regime, induced by strong, anisotropic exciton resonances. Using imaging spectroscopic ellipsometry (ISE), we characterize the complex dielectric function along in-plane and out-of-plane directions down to 190 nm (6.53 eV), revealing a type-II hyperbolic window in the DUV regime. This hyperbolicity supports hyperbolic exciton polaritons (HEP) with high directionality and slow group velocity. Our findings establish hBN as a promising platform for nanophotonic applications in the technologically significant DUV spectral range.

## Introduction

Nanoscale light manipulation is crucial for developing next-generation nano-optical devices for sensing[1], waveguiding[2], and imaging[3]. Hyperbolic media, characterized by permittivities of opposite sign along different crystal axes (e.g., $Re(\varepsilon_{\parallel}) \times Re(\varepsilon_{\perp}) < 0$), have attracted significant attention due to their ability to support large wavevectors, enabling extreme sub-diffraction light confinement and large photonic density of states (PDOS)[4,5]. Since hyperbolic dispersion requires large anisotropy, it is typically achieved in artificially made hyperbolic metamaterials (HMMs)[3,6-9], which are realized by subwavelength multilayer stacks of alternating dielectric and metallic films, exhibiting hyperbolic dispersion because it heavily relies on the plasmonic resonance of the metal layers[4]. These HMMs exhibit an open isofrequency surface, supporting the propagation of high-momentum waves and enabling unique phenomena like negative refraction[10], sub-diffraction imaging via hyperlensing[3,8,9], and substantial enhancement of the Purcell factor[11].

Van der Waals (vdW) two-dimensional (2D) materials exhibit inherent high anisotropy along the out-of-plane direction due to their layered structure. This pronounced anisotropy is a key prerequisite for hyperbolic dispersion[12-14]. Moreover, the pronounced anisotropy of certain vdW 2D crystals gives rise to strongly direction-dependent resonances—phonons, plasmons, and excitons—which profoundly impact their dielectric response[15]. Prior results have revealed hyperbolic behavior originating from the anisotropic resonances of phonons in α-$MoO_3$[16] and hexagonal boron nitride (hBN)[17], plasmons in $WTe_2$[18], and excitons in monolayer black phosphorus (BP)[19] and chromium sulfide bromide (CrSBr)[20] at cryogenic temperatures.

Natural hyperbolic media support hybrid electromagnetic modes (hyperbolic polaritons) that enable broadband, deeply sub-diffractional confinement with strongly directional propagation. Natural hyperbolic phonon polaritons were first observed in hBN crystals experimentally by near-field measurements in the mid-infrared (MIR)[21]. Since then, various hyperbolic polaritons driven by phonons, plasmons, and excitons have been demonstrated across many material platforms. Their extreme mode confinement, high PDOS, and directional energy flow promise new capabilities in sub-diffraction imaging, near-field focusing, and nanophotonic applications[15]. While phonon- and plasmon-mediated hyperbolicity has been extensively explored, exciton-induced hyperbolicity remains relatively rare, as it requires materials with exceptionally strong excitonic resonances with extreme anisotropy.

hBN is a well-known vdW polar dielectric exhibiting hyperbolic dispersion in the MIR regime due to anisotropic phonon resonances[15,21]. It has also attracted attention as an excitonic material due to its strong emission in the deep-ultraviolet (DUV) regime (around 180 ~ 300 nm, 4.13 ~ 6.89 eV)[22]. Even though the fundamental bandgap of hBN is an indirect bandgap (K→M point) in the DUV regime, as proved by the two-photon excitation spectroscopy[23], it also has a direct transition near ~ 6.1 eV at the K valley, supported by numerous density-functional theory (DFT) calculations and experiments[24-26]. Strong in-plane covalent bonding and weak out-of-plane van der Waals interactions create tightly bound 2D excitons with the large binding energy, $E_b$ ~700 meV, for the lowest exciton[24,27-29]. In addition, the electronic flatband of hBN in the Brillouin zone facilitates strong light-matter interaction[24].

We report the first observation of natural hyperbolic dispersion in hBN, arising from its anisotropic strong exciton resonances in the DUV regime. Using imaging spectroscopic ellipsometry (ISE), we characterize the complex refractive index and exciton resonance properties of hBN along in-plane (∥) and out-of-plane (⊥) directions. hBN shows pronounced exciton oscillator strength in the in-plane direction compared to the out-of-plane direction, creating a hyperbolic window in the DUV regime. This hyperbolicity enables a transition in the isofrequency contour from elliptic (closed) to hyperbolic (open) shape, supporting hyperbolic exciton polaritons (HEP) with high directionality and confinement[15,20]. Our findings enable light manipulation at the nanoscale in the DUV regime, which is critical for extreme-resolution photolithography using 193 nm ArF excimer lasers and DUV nonlinear optics.

## Results and Discussion

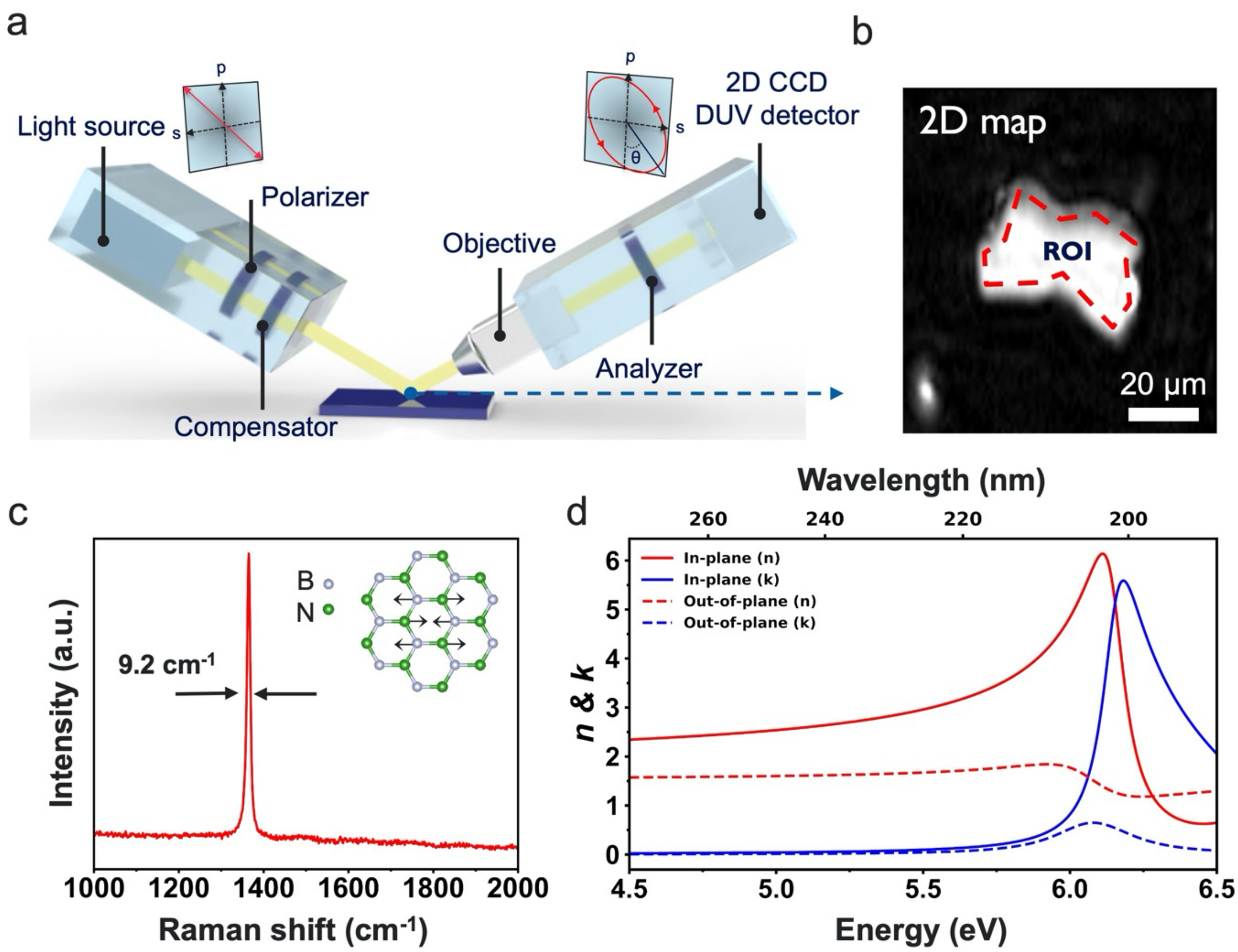

**Figure 1. Imaging spectroscopic ellipsometry measurement of hBN.** (a) Schematic illustration of ISE. Unlike conventional spectroscopic ellipsometry, imaging ellipsometry uses a high-resolution (~1 μm) 2D charge-coupled device (CCD) detector with an objective lens (7×) to map Ψ and Δ values, allowing the extraction of optical properties from even small flakes. (b) Example of the mapped image data of a small hBN flake using ISE. (c) Raman spectrum obtained from measured hBN shows a clear $E_{2g}$ resonance with a narrow FWHM, demonstrating the high quality of the measured hBN. (d) Refractive index (n) and extinction coefficient (k) of hBN along the different optical axes (in-plane (∥) and out-of-plane (⊥)) showing strong excitonic features of hBN. These extracted values from ISE are in agreement with prior measurements using synchrotron radiation and inelastic electron scattering[26,30].

While hBN's optical properties in the DUV have been widely studied, its electronic band structure calculation from the *ab initio* approach and its transitions by inelastic electron scattering (IES) measurement[24,26,28,31], there are very few spectroscopic-ellipsometry measurements that provide sensitive measurements[14,30]. Particularly, refractive indices resolved separately along the in-plane and out-of-plane directions are rare, yet they provide indispensable insight into the material's anisotropy and hyperbolic nature in the DUV excitonic regime. Recent ellipsometry studies have revealed a high refractive index and strong birefringence of hBN in the UV regime (around 250 nm), although they did not capture the excitonic resonance in the DUV regime[14]. We first characterize the excitonic features and refractive index of hBN along in-plane and out-of-plane directions using ISE down to the DUV to NIR regimes (190 ~ 1000 nm, 1.24 ~ 6.53 eV). First, we prepare the hBN sample using mechanical exfoliation with tape and transfer it onto the target substrate (See Methods). We chose the substrate $Al_2O_3$ which has a wide bandgap to increase the signal intensity and fidelity in DUV regime. When linearly polarized light is incident on a sample, strong excitonic resonances in its dielectric function cause variations in the Fresnel reflection coefficients $r_p$ and $r_s$, consequently changing the ellipsometry parameters $\Psi$ (the amplitude ratio) and $\Delta$ (the phase difference) with high sensitivity, enabling exciton behavior characterization[32]. Furthermore, obliquely incident light distinguishes a sample's electromagnetic response parallel to the surface from that perpendicular to it (Fig. 1a)[32]. As shown in Fig. 1b, an exfoliated hBN flake was characterized by ISE. The ~ 1 μm lateral resolution of our ISE allows detailed optical properties ($\Psi$ and $\Delta$) map data to be acquired as a function of incidence angle and wavelength on the small flakes that are inaccessible with conventional spectroscopic ellipsometry. The Raman spectrum of hBN exhibits a sharp peak at 1370 $cm^{-1}$ with a full width at half maximum (FWHM) of approximately 9.2 $cm^{-1}$, indicative of the material's high quality and the characteristic $E_{2g}$ vibrational mode, corresponding to the in-plane stretching of B and N atoms, as illustrated in the inset (Fig. 1c)[27]. This minimal structural disorder and low strain are essential for preserving strong excitonic resonances. We obtain the $\Psi$ and $\Delta$ (Supporting information (SI) Fig. S1) map and make the optical model using multiple Lorentz oscillator models to capture the exciton resonance behavior in hBN (see SI Note 1)[32]. Because hBN is uniaxial material, we modelled the in-plane and out-of-plane dielectric functions independently using distinct multiple Lorentz oscillator models. To improve the reliability of the results, multiple hBN thicknesses are measured at various incidence angles between 40° and 70° with respect to the sample's surface. The large variation in incidence angle, and thick flakes, yields a strong out-of-plane signal allowing us to determine the uniaxial dielectric function with high fidelity. The multiple Lorentz oscillator model captures the interband transitions in hBN along different optical axes with high accuracy, showing strong agreement with the experimentally obtained $\Psi$ and $\Delta$ values across various incident angles and hBN thicknesses (SI Fig. S2). The measured hBN's complex refractive index is shown in Fig. 1d along the in-plane (∥) and out-of-plane (⊥) directions. In the in-plane direction, hBN shows a strong exciton oscillation around 6.14 eV with a pronounced extinction coefficient ($k$), corresponding to the direct $\pi \rightarrow \pi^*$ transition at the K band in reciprocal space. This observation aligns well with the theoretical electronic band calculation using the *ab initio* (GW+BSE) method[24,28] as well as prior experimental measurements using synchrotron radiation[30] and inelastic electron scattering[26]. Although bulk hBN possesses an indirect bandgap, phonon-assisted (indirect) optical transitions are essentially absent from the measured spectra since their oscillator strength is orders of magnitude weaker than that of the dominant excitonic resonance[33,34]. Furthermore, the substantial refractive index (n ~ 2.5 at 250 nm) , while the loss remains low, due to strong excitons at the UV and DUV regime of hBN shows appliance to the UV optics primarily relied on the low refractive index materials such as fused silica ($SiO_2$, n ~1.5 at 250 nm)[35] and magnesium fluoride ($MgF_2$, n ~ 1.4 at 250 nm)[36]. On the other hand, hBN exhibits the interband transition along the out-of-plane direction with a small extinction coefficient (*k*) compared to that of the in-plane components due to its highly in-plane oriented exciton nature[27]. This observation is well aligned with the previous literature[26], manifesting strong and anisotropic excitonic behavior, which induces hyperbolic dispersion, discussed later.

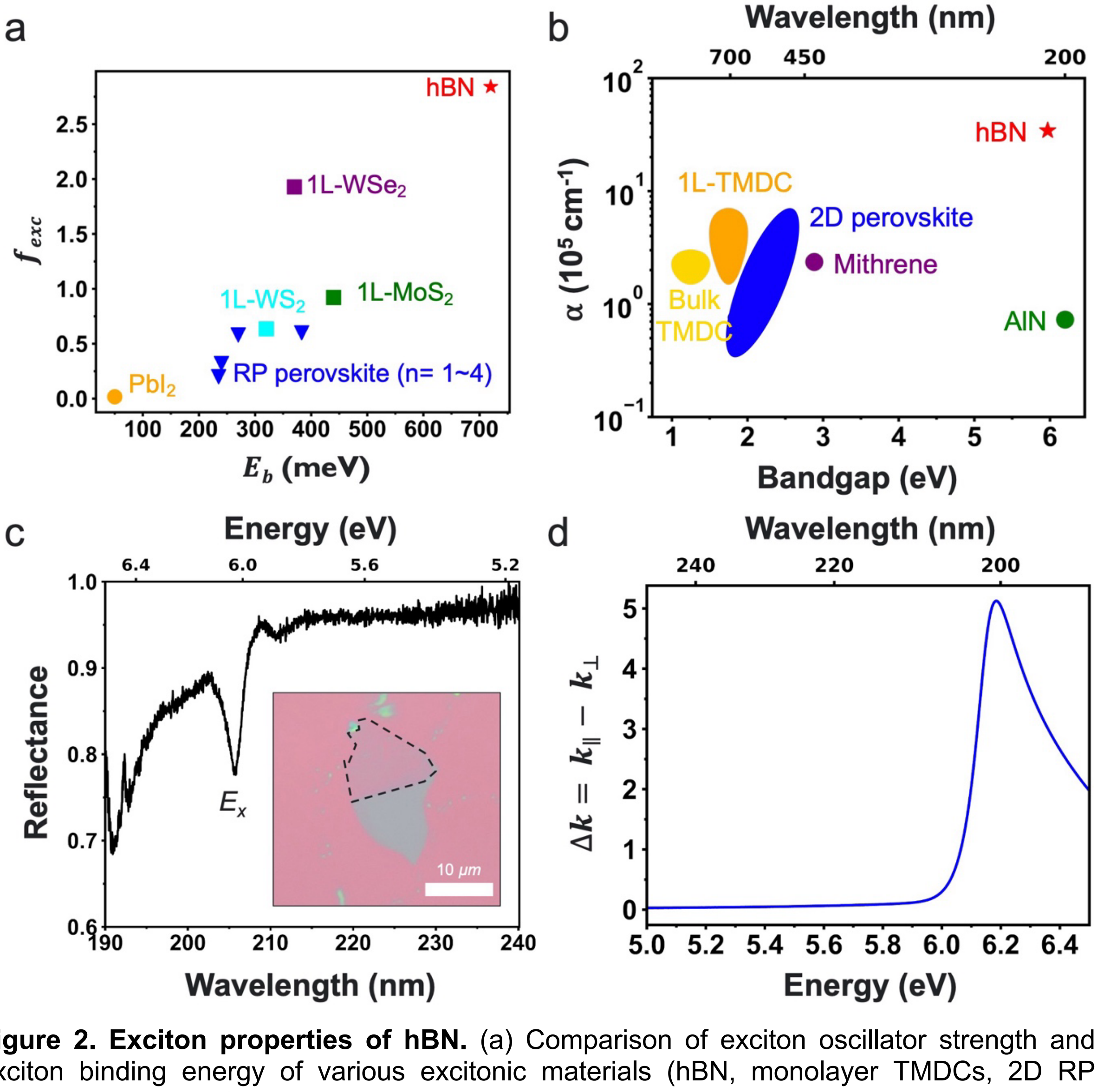

**Figure 2. Exciton properties of hBN.** (a) Comparison of exciton oscillator strength and exciton binding energy of various excitonic materials (hBN, monolayer TMDCs, 2D RP perovskite, and $PbI_2$), showing hBN's strong oscillator strength and exciton binding energy compared to the other excitonic materials. The exciton oscillator strengths are calculated using Eq. (1), and parameters such as interband oscillator strength, Bohr radius, and $E_b$ are adapted from prior literature[19,28,37-42]. (b) Comparison of the absorption coefficient of various excitonic (hBN, Bulk and monolayer of TMDCs ($MoS_2$, $MoSe_2$, $MoTe_2$, $WSe_2$, and $WS_2$) , 2D RP and DJ perovskites, mithrene, and AlN)[41,43-47], illustrating significant absorption coefficient due to the tightly bound excitons in hBN (c) Reflectance spectrum ($R_{hBN}/R_{Al}$) in a 5L-hBN/Al structure, demonstrating the high absorbance from the thin hBN layer due to strong exciton resonance. The reflectance dips around 190 nm (6.53 eV) are originated from strong

absorption of air. Inset shows image of thin hBN flake on a Si/$SiO_2$ substrate before the pick-up transfer. Note that the thickness of hBN was determined by the ISE measurement (SI Fig. S4) (d) Calculated linear dichroism ($\Delta k = k_{\parallel} - k_{\perp}$) as a function of photon energy, revealing a large dichroism originated from anisotropic exciton resonances.

Since the excitonic response critically governs the optical properties and potential hyperbolic behavior of 2D materials, we investigate the strong excitonic feature of hBN to analyze the origin of hyperbolicity by evaluating its exciton oscillator strength from the interband oscillator strength in ellipsometry measurements (Fig. 2a). Generally, because 2D materials strongly support confined excitons in two dimensions with reduced dielectric screening, they exhibit significantly larger $E_b$ and strong oscillator strengths compared to their bulk counterparts[48]. The exciton oscillator strengths ($f_{exc}$) can be calculated using the simple formulation proposed by *Ishihara et al.*[42].

$$f_{exc} = \frac{2fS_0}{\pi a_{2D}^2} \ or \ \frac{2fV_0}{\pi a_{3D}^2} \tag{1}$$

, where $f$ is the interband oscillator strength, $S_0$ ($V_0$) is a unit-cell area (volume), and $a_{2D\ (3D)}$ is the material's 2D (3D) Bohr radius. Since a smaller exciton Bohr radius is advantageous for the strong oscillator strength, $f_{exc}$ shows an inversely square (cubic) proportional relation to Bohr radius, respectively. Tightly bound hBN excitons exhibit a small Bohr radius[27], resulting in a pronounced $f_{exc}$ and $E_b$ compared to the well-known highly excitonic materials, such as monolayer transition metal dichalcogenides (TMDCs), and 2D Ruddlesden-Popper (RP) perovskites. Monolayer TMDCs exhibit a large oscillator strength, although it is slightly smaller than that of hBN. Consequently, several studies have reported hyperbolic dispersion of 2D vdW materials under specific conditions where excitonic resonances are enhanced, such as at low temperatures, highlighting the critical role of excitons in governing hyperbolicity[20,49]. Furthermore, RP perovskites exhibit a near-hyperbolic dispersion around exciton resonance, characterized by $Re(\varepsilon_{\parallel})$ approaching zero while $Re(\varepsilon_{\perp})$ remains positive, suggesting that true hyperbolicity could potentially emerge at low temperatures[41]. For comparison, $PbI_2$, which does not support strong excitonic effects, exhibits small values of $f_{exc}$ and $E_b$, indicating that excitons play an important role in the material's optical properties[42]. Fig. 2b shows the absorption coefficient, $\alpha = 4\pi k/\lambda$, of various well-known excitonic materials including hBN (Bulk and monolayer TMDCs, 2D RP and Dion-Jacobson (DJ) perovskites, mithrene (AgSePh), and AlN). The giant excitonic oscillator strength of hBN enables a strong absorption peak around the exciton energy approximately $3.5 \times 10^6\ cm^{-1}$ (SI Fig. S3), which is consistent with the observation in Fig. 2a. Because exciton oscillator strength scales with optical absorption, monolayer TMDCs, where reduced dielectric screening enhances excitonic features such as $E_b$ and $f_{exc}$ ,exhibit more significant absorption coefficients than their bulk counterparts. Notably, hBN's tightly bound exciton gives rise to absorption coefficients nearly an order of magnitude higher than other excitonic semiconductors. Even compared to wide-bandgap AlN, whose band edge lies similarly in the DUV, hBN's excitonic resonance boosts its absorption by approximately two orders of magnitude, underscoring the dramatic impact of its bound excitons. Therefore, exciton dominates the optical responses in the DUV regime of hBN.

To probe the strong excitonic response, we measure the reflectance spectrum of thin hBN (5 layers) on an Al substrate in the DUV regime. The spectrum reveals pronounced absorption near the exciton energy, as shown in Fig. 2c. Notably, most of the observed absorption originates from the hBN layer rather than the underlying Al substrate, as confirmed by the transfer matrix method (TMM)[50,51] calculation results (SI Fig. S5). More importantly, these strong excitons only exist along the in-plane direction, resulting in a large dichroism ($\Delta k = k_{\parallel} - k_{\perp}$) around 5, which reflects the difference in absorption properties along different optical axes, a prerequisite for natural hyperbolicity. Along with dichroism, hBN shows large birefringence in the DUV regime since dichroism and birefringence are related under the Kramers-Kronig relation (SI Fig. S6)[32]. In addition, the large birefringence of hBN, even well below the band gap where loss is low, makes it uniquely suited for UV and DUV polarization optics; an attribute few other materials can match[14].

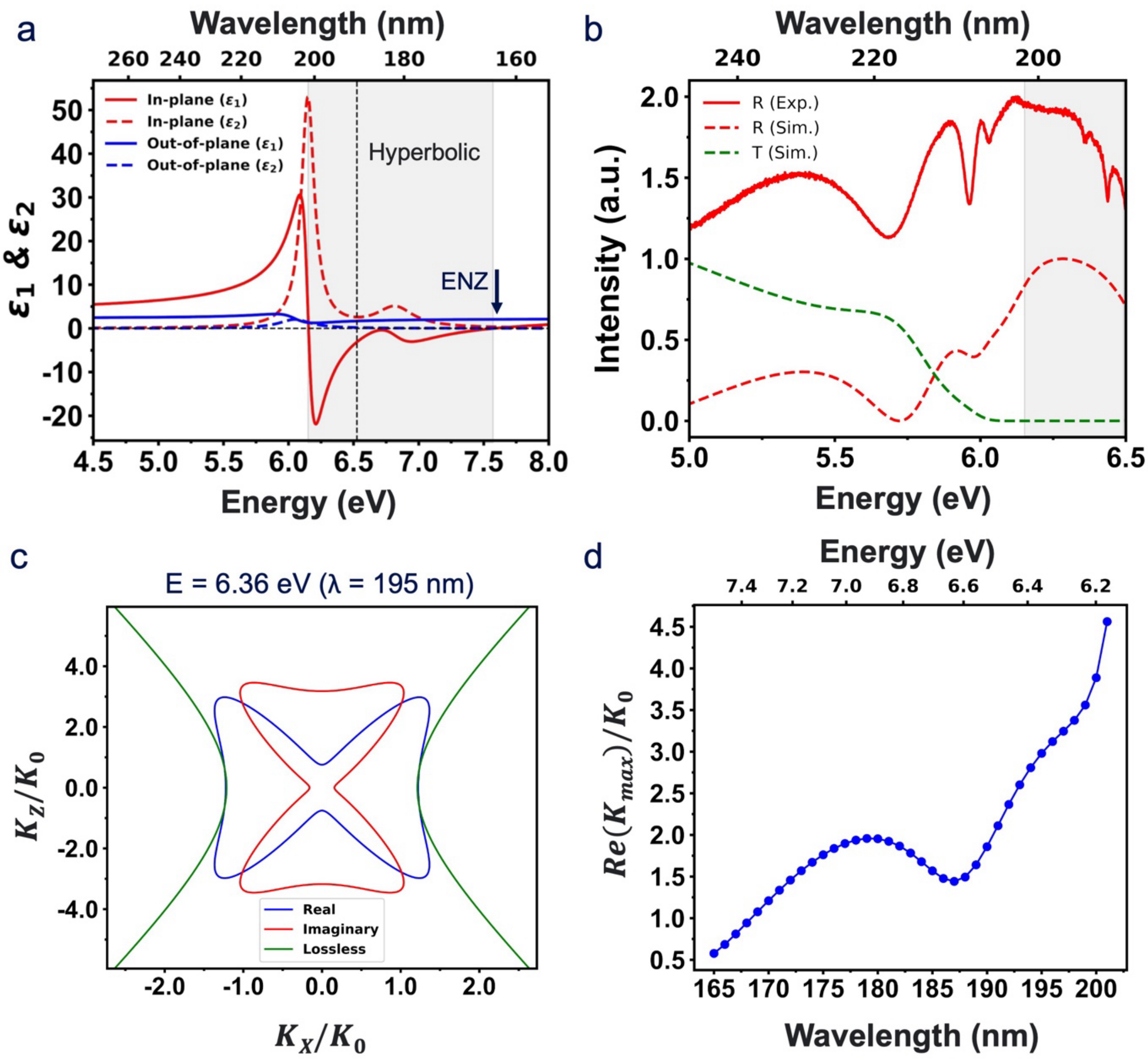

**Figure 3. Natural hyperbolic dispersion in hBN.** (a) Complex permittivities ($\varepsilon = \varepsilon_1 + i\varepsilon_2$) of hBN along different optical axes (in-plane (∥) and out-of-plane (⊥)) indicating apparent hyperbolic dispersion (grey shaded area) in the DUV regime with ENZ point. The permittivities are extrapolated based on the multiple Lorentz oscillator model obtained from the ISE measurement above (below) 6.53 eV (190 nm). Dashed vertical line represent the boundary

between experimental and extrapolated regime. (b) Measured (solid red curve) and simulated (dotted red curve) reflectance, along with simulated transmittance (dotted green curve), from a 63 nm hBN sample on an $Al_2O_3$ substrate in the DUV regime, showing the Reststrahlen (grey shaded area) with highly reflective properties. (c) Isofrequency surface of hBN at 6.36 eV (195 nm) for TM plane wave with real loss (red and blue curves) and no loss (green curve), showing clear hyperbolic dispersion arising from different signs of permittivities along different optical axes. (d) The max wavevector can be supported by the hBN as a function of wavelength (photon energy), reaching approximately ~4.6 around the exciton resonance.

As shown in Fig. 3a, we plot the complex permittivity ($\varepsilon = \varepsilon_1 + i\varepsilon_2$) of hBN, where $\varepsilon_1 = n^2 - k^2$ and $\varepsilon_2 = 2nk$, along both its in-plane and out-of-plane directions. The permittivity was directly retrieved from ISE measurements down to 190 nm (6.53 eV), a measurable wavelength regime from our tool. To extend our model to shorter wavelengths (~155 nm, ~8.0 eV), we extrapolated the data using a multi Lorentz oscillator model (See SI Note 1) obtained from multiple samples and incidence angles. Since the higher energy electron transition above 6.53 eV is out of range from our tool, we adapted the previous synchrotron study to capture the high energy interband transitions (e.g., $\pi \rightarrow \pi^*$ transition and $\sigma \rightarrow \sigma^*$ transition)[30]. Due to its pronounced anisotropic excitonic resonances, hBN supports a well-defined DUV type-II hyperbolic window around 6.17 eV ~ 7.56 eV (164 nm ~ 201 nm), where $\varepsilon_{1,\parallel} < 0$ and $\varepsilon_{1,\perp} > 0$. We further identify an intrinsic epsilon-near-zero (ENZ) wavelengths in hBN near 7.56 eV, where the losses are minimal. These naturally occurring hyperbolicity and ENZ points promise new opportunities for DUV nanophotonic device engineering and quantum-optical applications[52,53]. In its type-II hyperbolic regime, hBN exhibits a metallic in-plane permittivity while maintaining a dielectric response along the out-of-plane direction, a condition that underpins the formation of its Reststrahlen band. Our measured and simulated reflectance spectrum from hBN (63 nm) on $Al_2O_3$ substrate exhibit the high reflectivity plateau of the hBN's Reststrahlen band (grey-shaded region), an intrinsic consequence of its hyperbolic dispersion (Fig. 3b)[17,24]. The Reststrahlen band is observed in hBN flakes of various thicknesses (see SI Fig. S7a). In addition, the simulated transmission spectrum dips to near zero, confirming that electromagnetic waves cannot propagate normally through the hBN in this band except as evanescent/high-k modes. In the strong-response regime dominated by coherent radiative decay in the Reststrahlen band, hBN exhibits highly reflective behavior. However, as the thickness decreases and the hBN approaches the optically thin limit, a weak-response regime governed by damping emerges, where the broad Reststrahlen band transitions into a single resonance peak (SI Fig. S7b)[54]. The Reststrahlen band shows a highly reflective nature for the wide range of incident angles in the strong response regime (SI Fig. S7c).

The hyperbolic features are manifested only under transverse-magnetic (TM)-polarized light, as it probes both in-plane and out-of-plane dielectric responses. In contrast, transverse-electric (TE)-polarized light is sensitive only to the in-plane permittivity and may not exhibit hyperbolic behavior. For the TM plane wave, Maxwell's equation yields a dispersion relation: $\frac{k_x^2}{\varepsilon_{1,\perp}} + \frac{k_z^2}{\varepsilon_{1,\parallel}} = k_0^2$, where $k_0 = \frac{\omega}{c}$ is the free space wavevector, and $k_x$ and $k_z$ are the wavevectors in the x-plane and z-plane, respectively. Based on the dispersion relation, Fig. 3c shows the clear hyperbolic dispersion of hBN at 6.36 eV (195 nm) for each lossy (red and blue) and lossless (green) case, while hBN exhibits an elliptical dispersion outside the hyperbolic region as expected (SI Fig. S8). The anisotropic excitonic resonance gives rise to the hyperbolic dispersion, supporting highly confined light and large PDOS, enabling the formation of hyperbolic polaritons discussed later[15,20,55]. In an ideal, lossless hyperbolic medium, the dispersion relation admits modes with infinitely large wavevectors in theory, formally yielding an infinite PDOS. In practice, however, material absorption introduces damping that imposes a finite cutoff on the maximum supported k-vector, resulting in a closed isofrequency surface[56], as shown in Fig. 3c. Thus, it becomes essential to quantify the impact of absorption, especially in regimes where hyperbolic dispersion is driven by exciton resonances, since these same resonances impart strong intrinsic damping. Over hBN's hyperbolic band, we compute the largest wavevector the material can support as a function of photon energy (wavelength). Approaching the exciton resonance, hBN's hyperbolic bands extend to larger wavevectors, albeit with rising absorption. SI Fig. S9 compares the loss-inclusive isofrequency contours at multiple photon energies (wavelength): as one approaches the exciton resonance (～6.14 eV), the real isofrequency surface expands (indicating larger wavevector) while the imaginary isofrequency surface grows simultaneously, indicating increased damping near the resonance. In its most extreme hyperbolic regime, hBN reaches $\mathrm{Re}(K_{max})/K_0$ ~ 4.6 (Fig. 3d). Additionally, the enlarged wavevector space within hBN's hyperbolic band significantly enhances the PDOS, which is calculated by the Green's functions, resulting in an enhanced spontaneous emission rate, as quantified by the Purcell factor shown in the SI Fig. S10 [4,57].

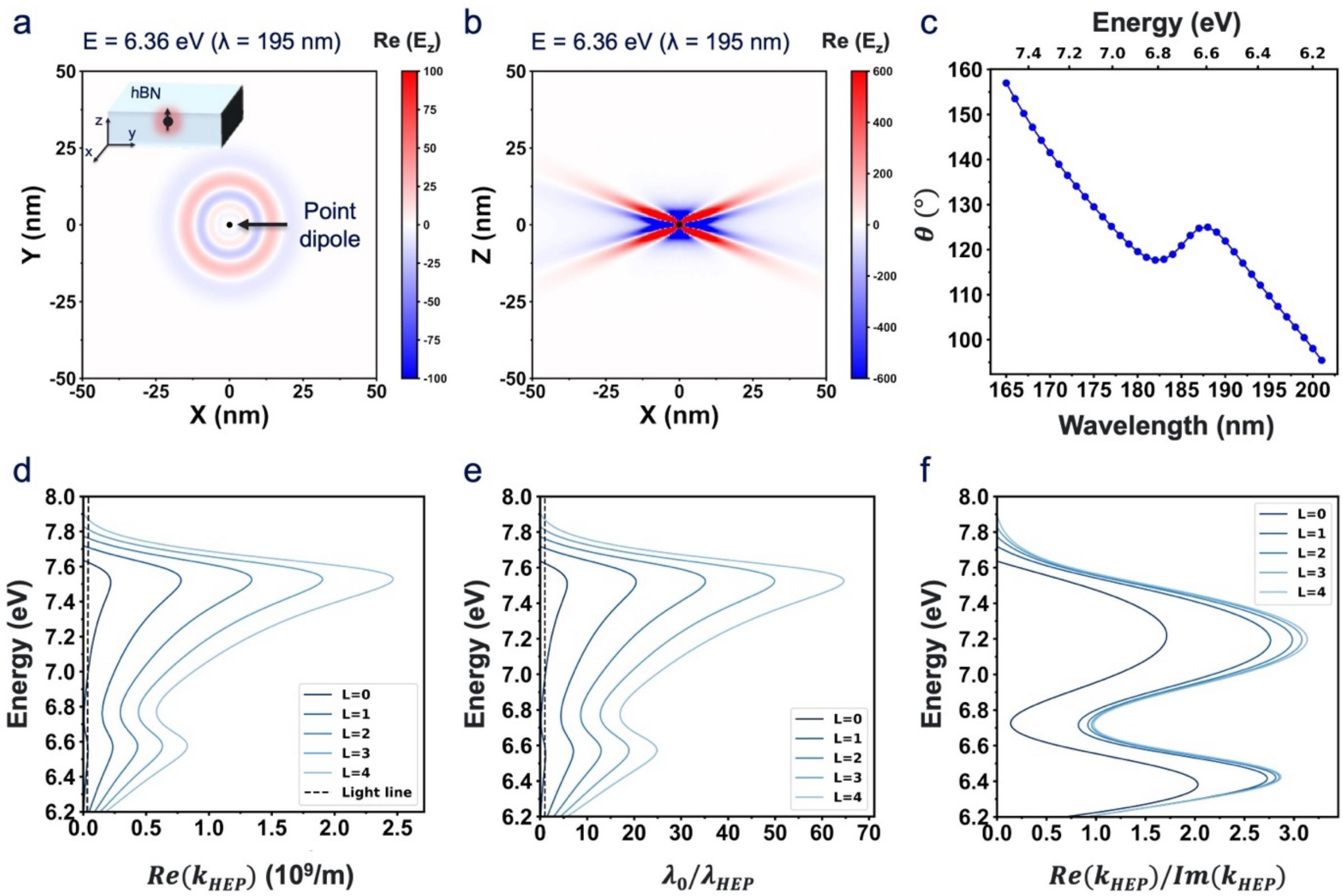

**Figure 4. HEP propagation in hBN slab.** Numerically simulated near-field wave propagation at 6.36 eV (195 nm) using finite-difference time-domain (FDTD) simulations, launched by a point dipole oriented along the Z-axis, is shown in (a) the in-plane (XY) direction, revealing isotropic and radial propagation, and (b) the out-of-plane (XZ) direction, exhibiting the highly directional wave propagation due to the hyperbolic dispersion. (c) Wave propagation angle as a function of photon energy (wavelength). (d) The HEP's analytical dispersion relation as a function of an in-plane wavevector up to L=4 mode, where $d_{hBN}$ is 10 nm, showing the highly confined nature of HEP. Calculated figure of merits: (e) Confinement factor $\lambda_0/\lambda_{HEP}$ and (f) Propagation loss properties ($Re(k_{HEP})/Im(k_{HEP})$) of HEP as a function of incident photon energy, showing highly confined propagation characteristic.

We further investigate the highly confined HEP originating from hyperbolic dispersion. The numerically simulated wave launched by a point source oriented along the z-axis, as illustrated in the inset of Fig. 4a, displays a circularly shaped radial wave propagation at 6.36 eV (195 nm) in the XY plane. Since radiation in the XY plane consists solely of the $E_z$ component, governed exclusively by a positive $\varepsilon_\perp$, the resulting radiation pattern exhibits the symmetric propagation shown in Fig. 4a. In contrast, the radiation pattern in the XZ plane comprises two components ($E_x$ and $E_z$), allowing the field to probe the hyperbolic dispersion. Consequently, a point source excitation in a type-II hyperbolic medium produces a sharply defined hyperbolic radiating pattern from the origin in the XZ planes (Fig. 4b), which is consistent with the isofrequency surface at 6.36 eV (195 nm) in reciprocal space in Fig. 3c. On the other hand, the point source launched at 4.96 eV (250 nm), which is outside of the hyperbolic window, does not show directional propagation as can be expected from its isofrequency contour (SI Fig. S11). Since HEP supports large momentum, they are difficult to excite via direct free-space incidence in the far field[1]. Thus, using a high-index prism such as the Otto and Kretschmann configuration under TM-polarized illumination enables efficient coupling to these high-k modes (SI Fig. S12)[58].

This hyperbolic dispersion enables highly directional propagation and high-electric field confinement of HEP in the near field, demonstrating high tunability of the wave on the nanoscale, in contrast to elliptical dispersion materials where energy spreads across the plane[15,59]. In addition, HEP exhibits directional propagation depending on the frequency, and the propagation angle (θ), the angle between the Poynting vector and the z axis, can be approximately calculated by Eq. (2)[60]:

$$\theta(\omega) = \pi/2 - \tan^{-1}(\sqrt{\varepsilon_\perp(\omega)}/i\sqrt{\varepsilon_\parallel(\omega)}) \tag{2}$$

In this regime, polaritons can only propagate at specific angles determined by the anisotropy of the material. The propagation angle (θ) decreases from 157° to 95° (Δθ ≅ 62°) within the hyperbolic window as photon energy decrease (wavelength increases) (Fig. 4c). Simulated wave propagation in the XZ plane as a function of photon energy exhibits pronounced directionality, in agreement with the calculations shown in Fig. 4c. (SI Fig. S13). We analytically solve the HEP dispersion relation for TM modes, in the thin hBN slab following the general equations for the TM polariton mode given by Eq. (3)[61]:

$$k_{HEP} = \frac{i}{d_{hBN}}\sqrt{\frac{\varepsilon_\perp}{\varepsilon_\parallel}}[2\tan^{-1}\left(i\frac{1}{\sqrt{\varepsilon_\perp\varepsilon_\parallel}}\right) + \pi L] \tag{3}$$

, where $k_{HEP}$ is the x component of the wavenumber, $d_{hBN}$ is the thickness of the hBN slab and L = 0, 1, 2 … are integers indicating the higher order mode. The analytical dispersion relation derived from Eq. (3) clearly reveals the HEP mode, which carries a larger momentum than that of a free-space photon, where the dashed line represents the light line (Fig. 4d). As expected for the hyperbolic polariton, the HEP's dispersion relation exhibits an increasing momentum as mode order increases. In addition, the group velocity ($v_g = \partial\omega/\partial k_{HEP}$), the slope of the dispersion curve, decreases with increasing momentum, reflecting the slow group velocity and strong spatial confinement of higher-order modes[17]. We extract the commonly used figure of merits for the polariton: the confinement factor ($\lambda_0/\lambda_{HEP}$) ,where $\lambda_0$ is the free-space wavelength, and propagation figure of merit of HEP, defined as $Re(k_{HEP})/Im(k_{HEP})$[55,62]. Fig. 4e shows the confinement factor depending on the mode order, showing the highly confined nature ($\lambda_0/\lambda_{HEP} > 1$) of HEP with an increase in confinement as the mode order increases. For the L=0 mode, the confinement factor reaches approximately 5.7, which is comparable to the values observed in exciton-polaritons in $WSe_2$[63]. In addition, the confinement factor reaches values exceeding 10 for higher-order modes, underscoring the extreme subwavelength confinement enabled by the hyperbolic dispersion in hBN. As shown in Fig. 4f, higher propagation figure of merit values indicates reduced relative propagation losses of HEP. This suggests that, despite their inherently large momentum, HEP can still propagate appreciable distance (~20 nm) before significant attenuation occurs[49,61]. Furthermore, the value of propagation figure of merit increases with mode order, reflecting improved loss-performance characteristics in higher-order modes. The calculated propagation length along the direction corresponding to the maximum real part of the wavevector is approximately 20 nm (SI Fig. S14)[56].

## Conclusion

We report the emergence of natural hyperbolic dispersion in hBN arising from strong and anisotropic excitonic resonances in the DUV regime. Using ISE with micron resolution, we extract the complex dielectric functions along both in-plane and out-of-plane directions, revealing a clear type-II hyperbolic window induced by excitonic anisotropy. We identify the Reststrahlen band originating from hyperbolicity through reflectance spectroscopy, showing good agreement with numerical calculations. Moreover, this hyperbolic dispersion supports highly confined polaritonic modes with large momenta, slow group velocities, and enhanced photonic density of states. We confirm this through numerical simulations and analytical modeling of the HEP dispersion relation, showing strong confinement factors and moderate propagation figures of merit. These findings establish hBN as a promising candidate for DUV nanophotonics, providing a pathway toward next-generation polaritonic devices for sub-diffraction imaging, DUV lithography, and quantum optical applications.

## Methods

### Sample preparation

The exfoliated hBN sample has prepared using two methods, direct transfer and pick-up transfer. For the direct transfer, the bulk hBN was exfoliated using blue tape and then directly dry transferred on top of the target substrate. For the pick-up transfer, mechanically exfoliated hBN flakes were picked up using a dry transfer method[64] with polycarbonate (PC) /polydimethylsiloxane (PDMS) and subsequently transferred onto the target substrate. After the transfer, the samples were immersed in chloroform for 1 hour to dissolve the PC film.

### Imaging spectroscopic ellipsometry

Imaging ellipsometry measurements for determining the dielectric function were performed using the Accurion EP4 system (Park Systems) over the 190–1000 nm spectral range. A 7× objective lens was used for measurements in the DUV regime. Multi-angle incidence measurements were conducted over an angular range of 40° to 70°. The measured mapping data were analyzed using the EP4 model and DataStudio software provided by the EP4 system.

### Raman characterization

Raman spectra were acquired using a Horiba LabRam HR Evolution confocal microscope, with a 600 grooves/mm grating and a 633 nm continuous-wave (CW) laser for excitation. The signal was detected using a CCD detector.

### Reflectance measurements

Reflectance measurements were performed using a Hamamatsu deuterium lamp as the light source. The DUV light was guided into a home-built confocal microscopy set-up, which was constructed using DUV-enhanced aluminum mirrors, a DUV-enhanced reflective objective with a numerical aperture of 0.4, $CaF_2$ lenses, and a $CaF_2$ beam splitter. A 4f optical system was implemented, and a size-tunable aperture was placed at the image plane to spatially filter the reflected signal, thereby isolating the reflectance from the hexagonal boron nitride samples while blocking contributions from the underlying substrate. The reflected light was guided into a spectrometer and detected using an electrically cooled Si CCD camera.

### Optical simulations

Electric field profiles in the near field were calculated using FDTD simulations performed with the commercial software Lumerical. Theoretical reflectance and absorbance are computed using the TMM with homemade Python code[50,51]. The refractive index of hBN was determined through imaging ellipsometry, while those of $Al_2O_3$[65] and Al[66] were adopted from previously reported literature values.

### Author Contribution

D.J. supervised and acquired funding for the project. D.J. and B.C. conceived and designed the experiment. B.C. performed the ellipsometry measurements and conducted the data fitting. B.C. performed the optical simulations and calculations in discussion with D.J., N.E., and J.L. W.C. fit preliminary ellipsometry data. K.Y. fabricated samples. S. J. and H.C. performed the reflectance measurement under the supervision of J. K. B.C. and D.J. wrote the manuscript with inputs from all authors. All authors discussed the results and revised the manuscript

### Acknowledgement

D. J., B.C., and J.L. acknowledge support from the Office of Naval Research Young Investigator Award, Metamaterials Program (N00014-23-1-203). This work was carried out in part at the Singh Center for Nanotechnology, which is supported by the NSF National Nanotechnology Coordinated Infrastructure Program under Grant NNCI-2025608. J.K. acknowledges the support from the Institute for Basic Science (IBS), Korea under Project Code IBS-R014-A1 and the National Research Foundation of Korea grants (NRF-2023R1A2C2007998). N. E. acknowledges partial support from the US Air Force Office of Scientific Research (AFOSR) Multidisciplinary University Research Initiatives (MURI) program with grant # FA9550-21-1-0312 and AFOSR grant # FA9550-23-1-030.7

Supporting Information

# Natural Hyperbolicity of Hexagonal Boron Nitride in the Deep Ultraviolet

Bongjun Choi[1], Jason Lynch[1], Wangleong Chen[2], Seong-Joon Jeon[3], Hyungseob Cho[3], Kyungmin Yang[1], Jonghwan Kim[3,4,5]*, Nader Engheta[1,2,6,7]*, Deep Jariwala[1,2]*

[1]Department of Electrical and Systems Engineering, University of Pennsylvania, Philadelphia, Pennsylvania 19104, United States

[2]Department of Materials Science and Engineering, University of Pennsylvania, Philadelphia, Pennsylvania 19104, United States

[3]Department of Materials Science and Engineering, Pohang University of Science and Technology, Pohang, Republic of Korea.

[4]Department of Physics, Pohang University of Science and Technology, Pohang, Republic of Korea.

[5]Center for van der Waals Quantum Solids, Institute for Basic Science (IBS), Pohang, Republic of Korea

[6]Department of Bioengineering, University of Pennsylvania, Philadelphia, Pennsylvania 19104, United States

[7]Department of Physics and Astronomy, University of Pennsylvania, Philadelphia, Pennsylvania, 19104, United States

* Corresponding authors: dmj@seas.upenn.edu, engheta@seas.upenn.edu, jonghwankim@postech.ac.kr

**The supporting information file includes:**

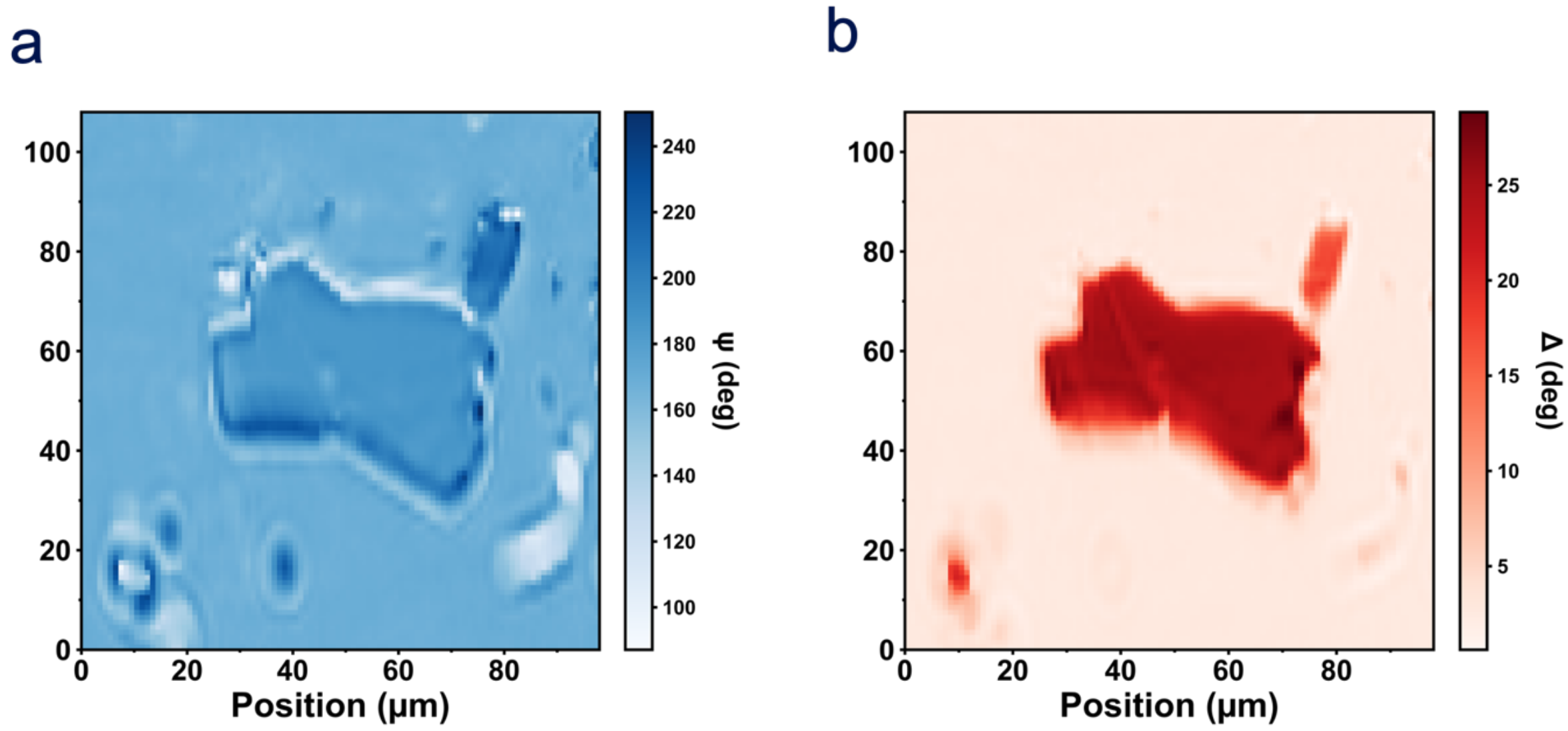

**Figure S1.** Two-dimensional (2D) map data at 250 nm of (a) *Ψ* (the amplitude ratio) and (b) *Δ* (the phase difference) value from hBN using imaging spectroscopic ellipsometry (ISE). ISE allows simultaneous measurement of all pixels inside the area of interest (AOI). Each pixel ~ 1 x 1 μm contains the *Ψ* and *Δ* information.

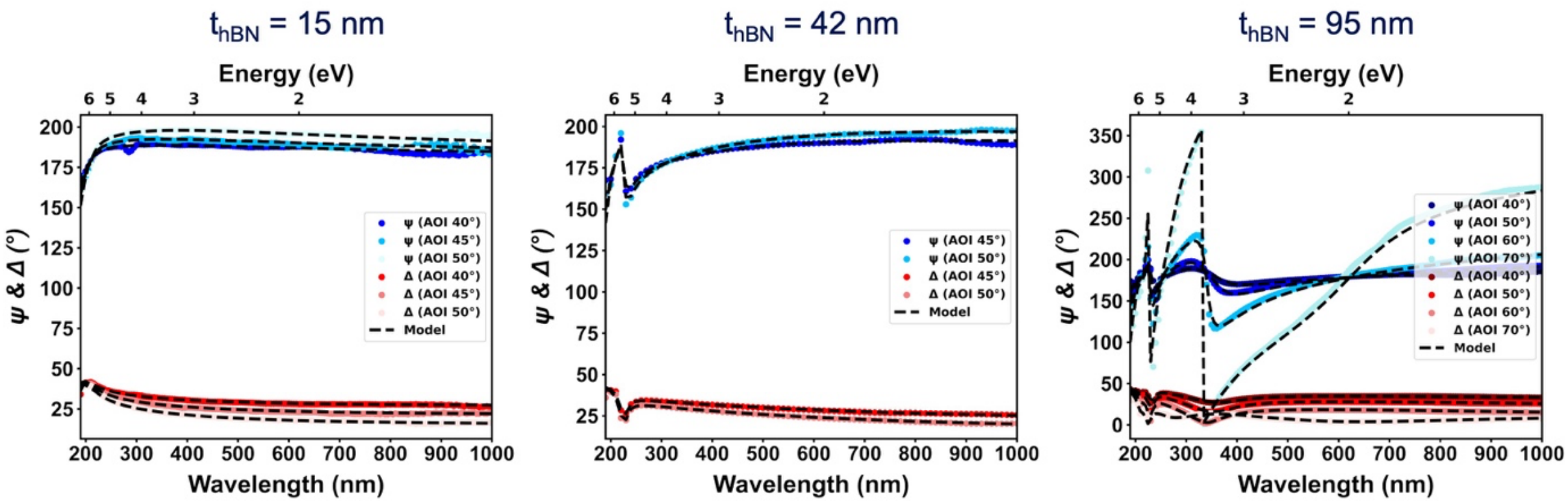

**Figure S2.** Measured *Ψ* and *Δ* values for hBN films with varying thicknesses (15, 42, and 95 nm) and incident angles (40°~70°), overlaid with fitted curves based on a multi-Lorentz oscillator model. hBN was treated as a uniaxial material with independently fitted in-plane and out-of-plane components[1], demonstrating high fidelity of the optical model.

**Note S1. Analysis of imaging spectroscopic ellipsometry data**

In ellipsometry measurement, the p- (transverse magnetic (TM)) and s- (transverse electric (TE)) polarized light are incident on the sample, measuring the polarization state of reflected light. Once the p- and s-polarizations of light hit the sample, each polarization of light shows different changes in amplitude and phase, and ellipsometry measures two parameters, $\Psi$ and $\Delta$, which represent the difference of amplitude and phase given by Eq. (1), where $r_p$ and $r_s$ are the Fresnel coefficients[1].

$$\rho = \frac{r_p}{r_s} = \tan\psi \, exp(i\Delta) \quad (1)$$

Ellipsometry is a model-based technique that requires two models to analyze the measured ISE data: the geometric structure of the sample and the dielectric function model. The geometric structure refers to the stacking sequence of the sample, including layer thicknesses. In order to increase the signal to noise ratio (SNR) in the DUV regime, we choose $Al_2O_3$ as a substrate due to its wide bandgap, and transferred the exfoliated hBN crystals. Thus, geometric structure is defined as an air / hBN (t nm) / $Al_2O_3$ (Substrate). Dielectric function model can describe the optical response, and especially electron transitions and exciton resonances are widely modelled using Lorentz oscillator model given by Eq. (2), where the resonance energy is $E_0$, $f$ is the oscillator strength, and a damping factor is $\Gamma$.

$$\varepsilon_{Lorentz}(E) = \frac{f E_0 \Gamma}{E_0^2 - E^2 - i\Gamma \mathrm{E}} \quad (2)$$

If there are multiple oscillators in the systems, the system can be described by the sum of Lorentz oscillator model described in Eq. (3), where $\varepsilon_\infty$ is a background permittivity accounting for high-energy electronic transitions that lie outside the measurable spectral range.

$$\varepsilon(E) = \varepsilon_\infty + \sum_i \varepsilon_{Lorentz}^i(E) \quad (3)$$

Since hBN has an anisotropy along z axis (uniaxial material), the dielectric tensor of hBN is represented by Eq. (4) since $\varepsilon_{xx} = \varepsilon_{yy}$.

$$\varepsilon = \begin{bmatrix} \varepsilon_{xx} & 0 & 0 \\ 0 & \varepsilon_{xx} & 0 \\ 0 & 0 & \varepsilon_{zz} \end{bmatrix} \quad (4)$$

Thus, $\varepsilon_{xx}$ (in-plane permittivity) and $\varepsilon_{zz}$ (out-of-plane permittivity) should be modelled respectively. Following previous literature[2-4], we model both in-plane and out-of-plane permittivity using multiple Lorentz oscillators within a uniaxial dielectric model, which accurately reproduces the experimental data, as shown in Fig. S2.

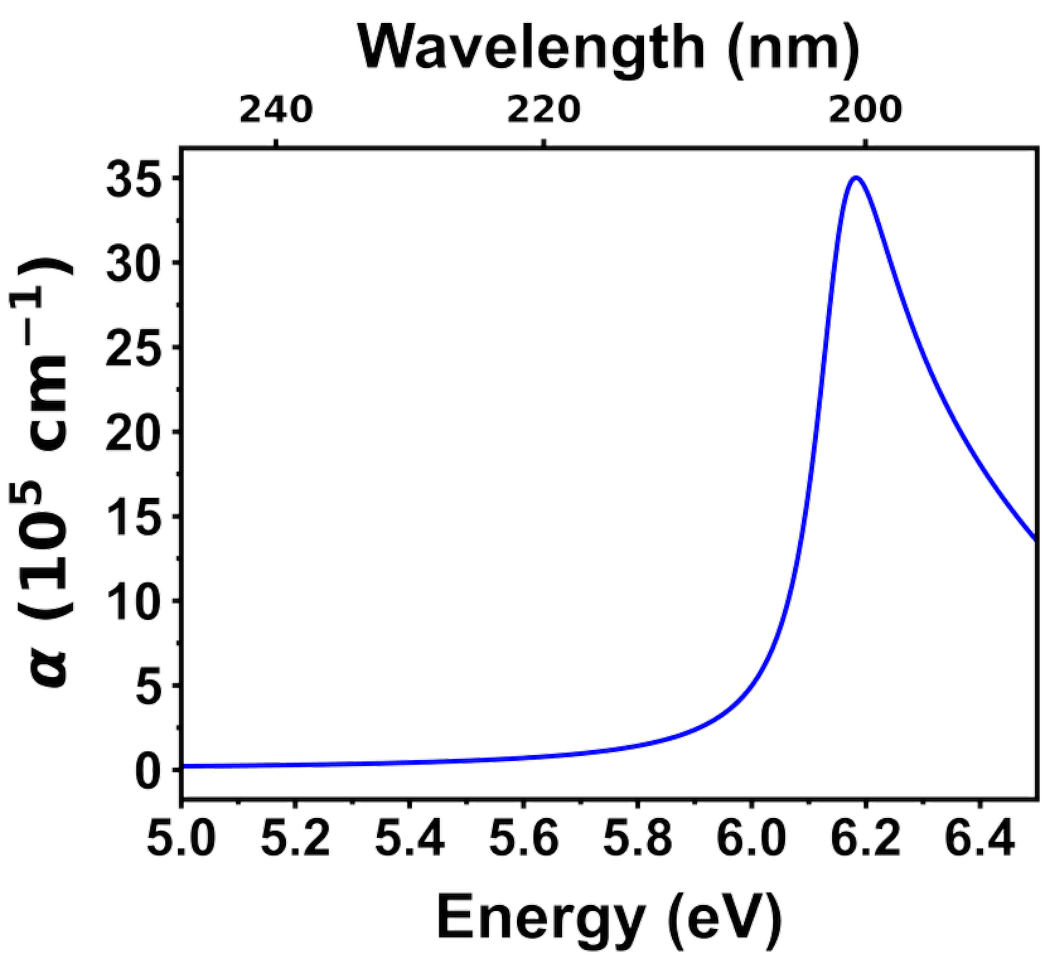

**Figure S3.** Absorption coefficient ($\alpha = 4\pi k/\lambda$) of hBN as a function of photon energy (wavelength), showing significant absorption coefficient due to the strong exciton oscillator strength.

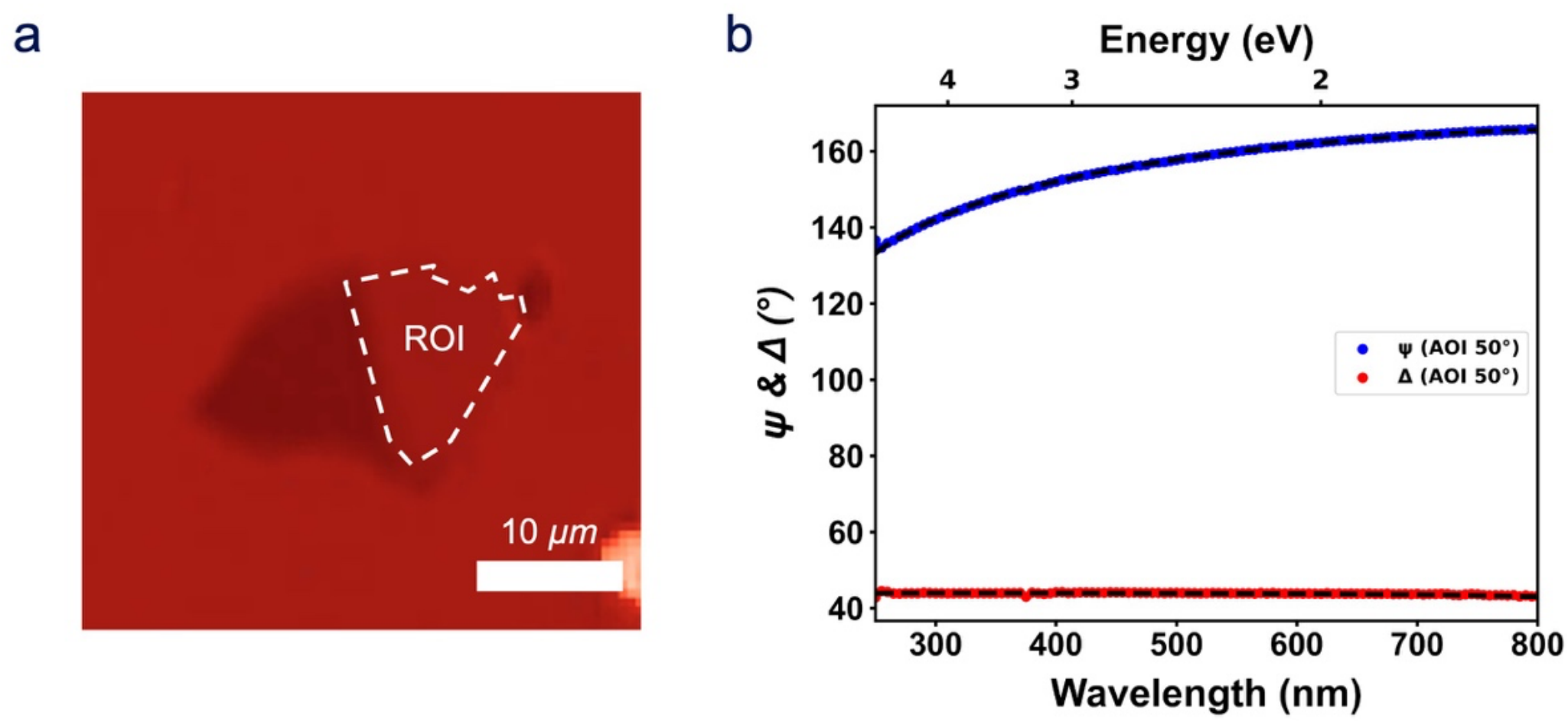

**Figure S4.** Determination of hBN thickness using imaging spectroscopic ellipsometry (ISE). (a) Mapped Δ values from thin hBN. (b) Fitted curves based on hBN refractive indices, indicating a hBN thickness of ~1.67 nm (5 layers).

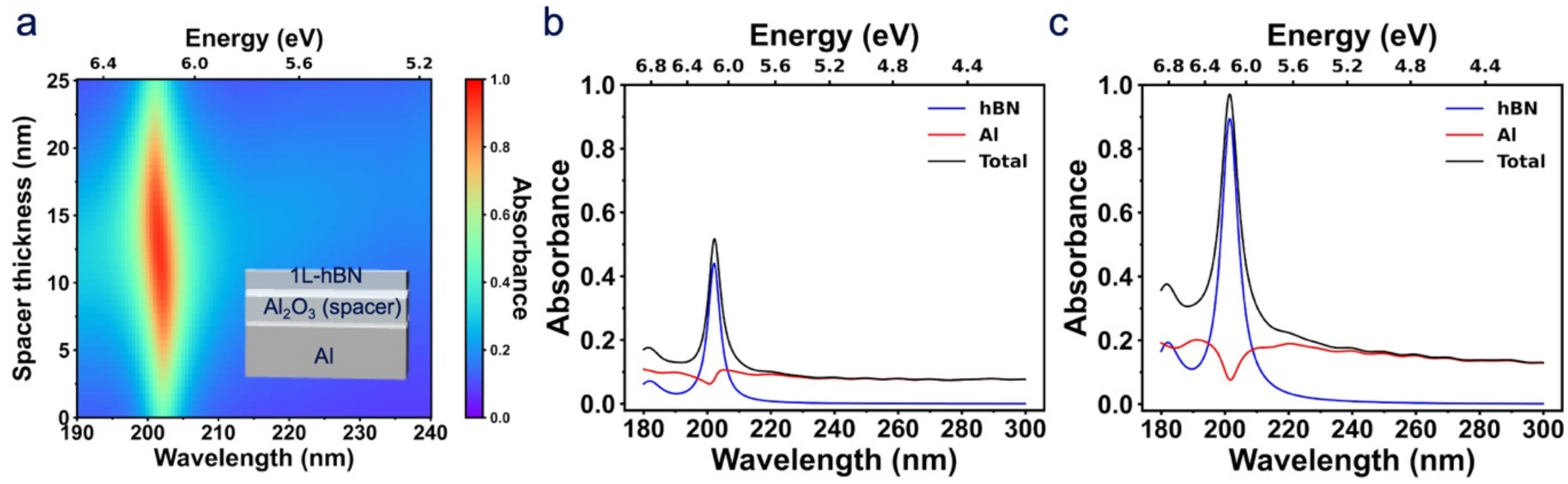

**Figure S5.** (a) Simulated absorbance spectrum using transfer matrix method (TMM)[5-7]calculation in a 1L-hBN/$Al_2O_3$/Al heterostructure as a function of $Al_2O_3$ spacer thickness, demonstrating the high absorbance from only a monolayer hBN layer due to strong exciton resonance. Note that we assume that monolayer hBN has the same refractive index as that of bulk for the calculation. Layer resolved absorbance spectrum depending on the presence and absence of the spacer structure: (b) 1L-hBN/Al (100 nm) and (c) 1L-hBN/$Al_2O_3$ (13.5 nm) /Al (100 nm). Even a monolayer of hBN on the Al substrate can absorb a significant amount of light. In addition, once an appropriate thickness of $Al_2O_3$ is added, absorbance can reach approximately unity absorption from a monolayer of hBN. Layer resolved absorbance calculation shows that this significant amount of absorption mostly comes from the hBN monolayer for both (b) and (c) structures, indicating the strong exciton strength in hBN.

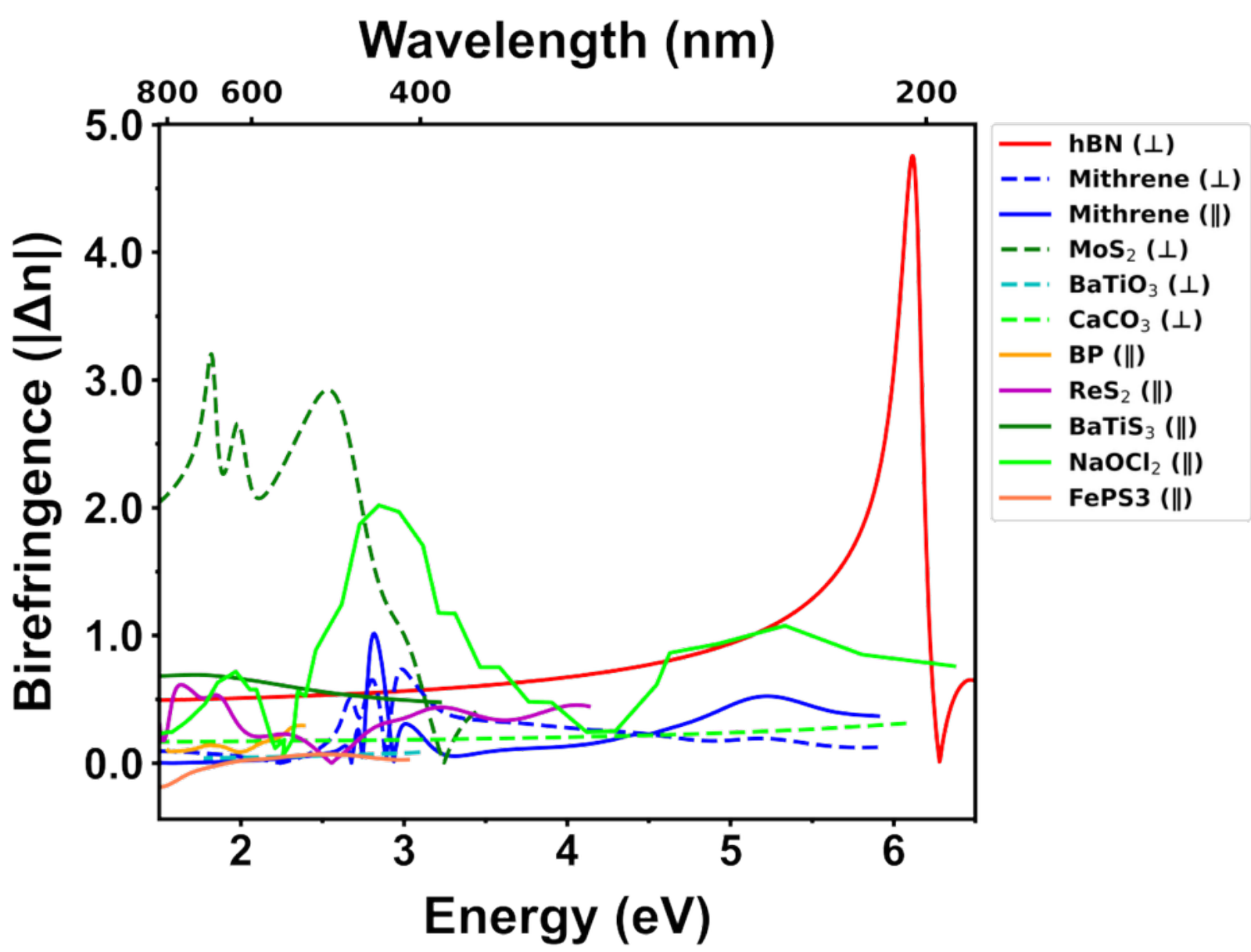

**Figure S6.** Comparison of the in-plane and out-of-plane birefringence magnitude as a function of wavelength for various birefringent crystals[8-16], including hBN. hBN exhibits exceptionally large birefringence spanning the visible to the deep ultraviolet (DUV) range. Notably, it shows significant birefringence in the DUV regime, surpassing recently reported $NaOCl_2$, an uncommon and remarkable property.

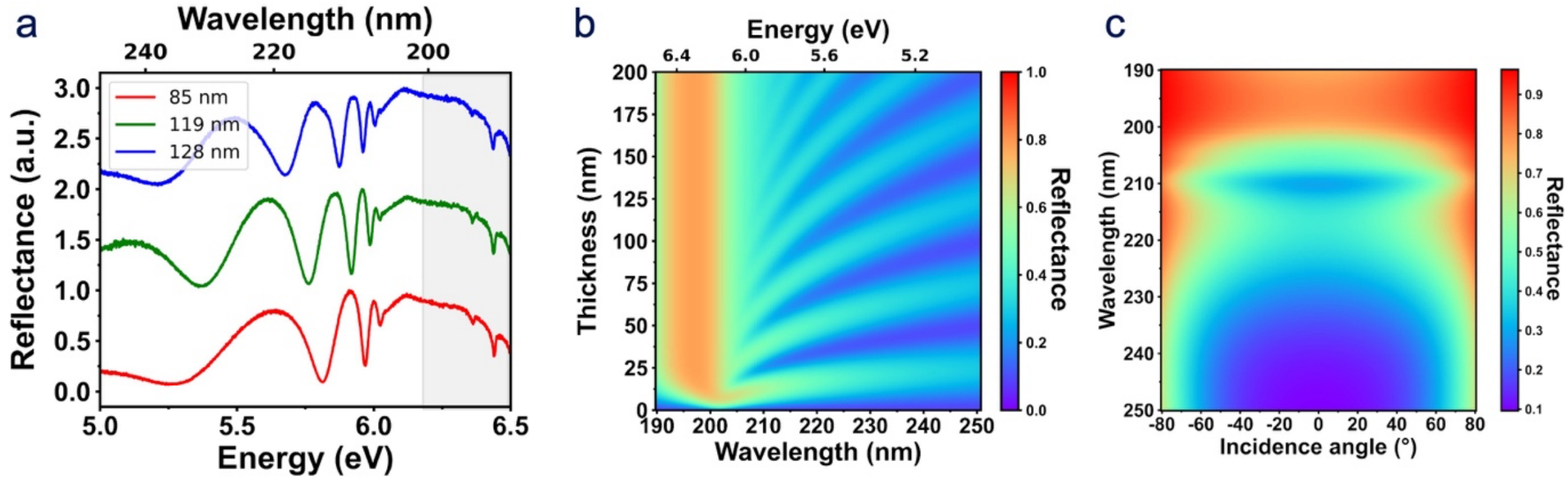

**Figure S7.** (a) The measured reflectance spectrum with hBN on $Al_2O_3$ substrate reveals the Reststrahlen band (grey-shaded region), exhibiting high reflectance across multiple hBN thicknesses. Simulated reflectance spectrum using TMM as a function of wavelength /thickness/incidence angle, showing the Reststrahlen band of hBN in the DUV regime in good agreement with experimental results in (a). (b) The thickness of hBN was changed on the $Al_2O_3$ substrate. As the hBN thickness approaches the optically thin limit, the Reststrahlen band changes to a single resonance. On the other hand, the thick enough hBN (> 25 nm), which is in the strong response regime, shows high reflectance. (c) Calculated reflectance as a function of the incidence angle of hBN (50nm) on the $Al_2O_3$ substrate structure, indicating a highly reflective nature of the Reststrahlen band in a wide range of incidence angles.

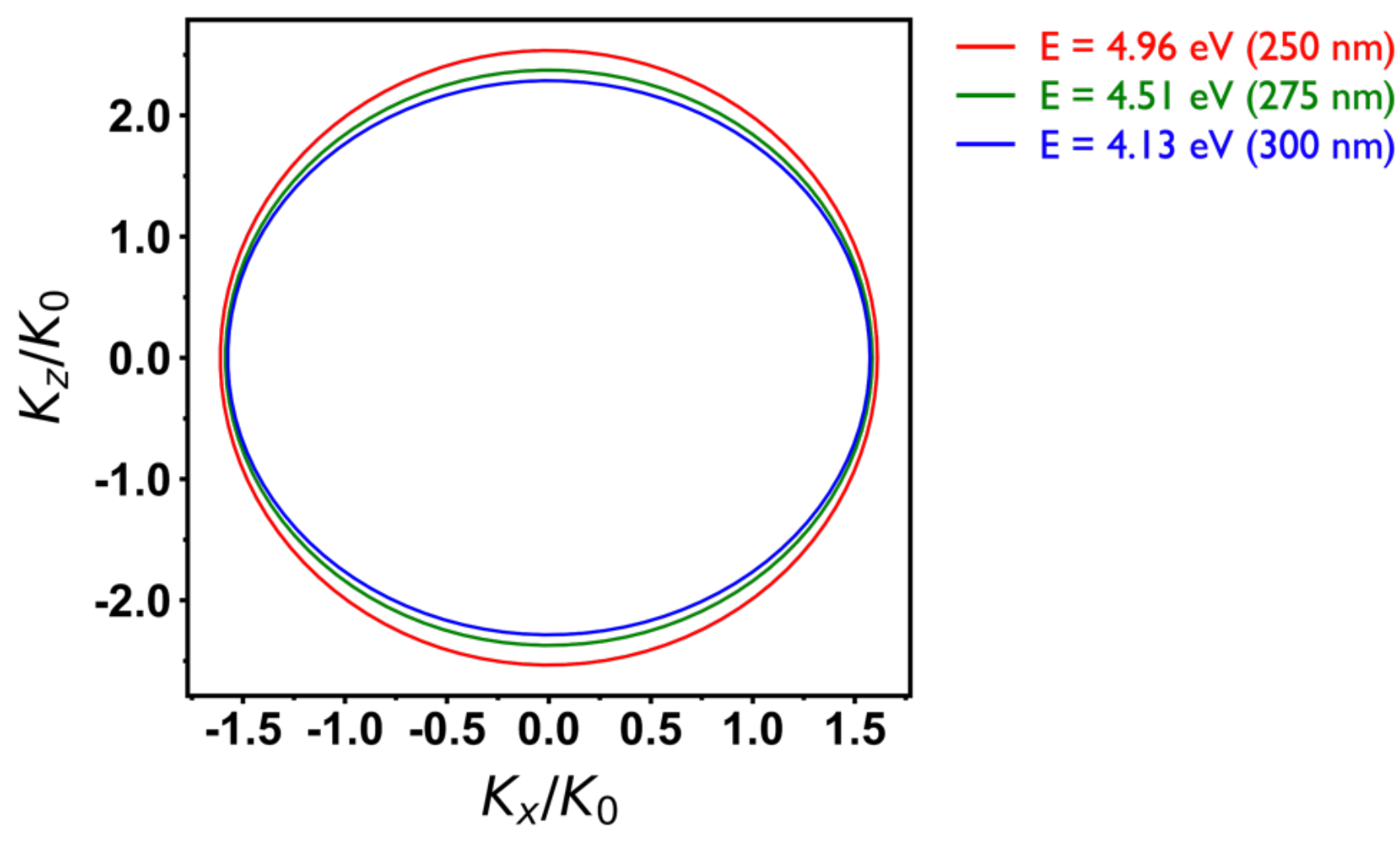

**Figure S8**. Isofrequency surface in various photon energies (wavelengths) of hBN outside of the hyperbolic regime, showing the elliptical dispersion.

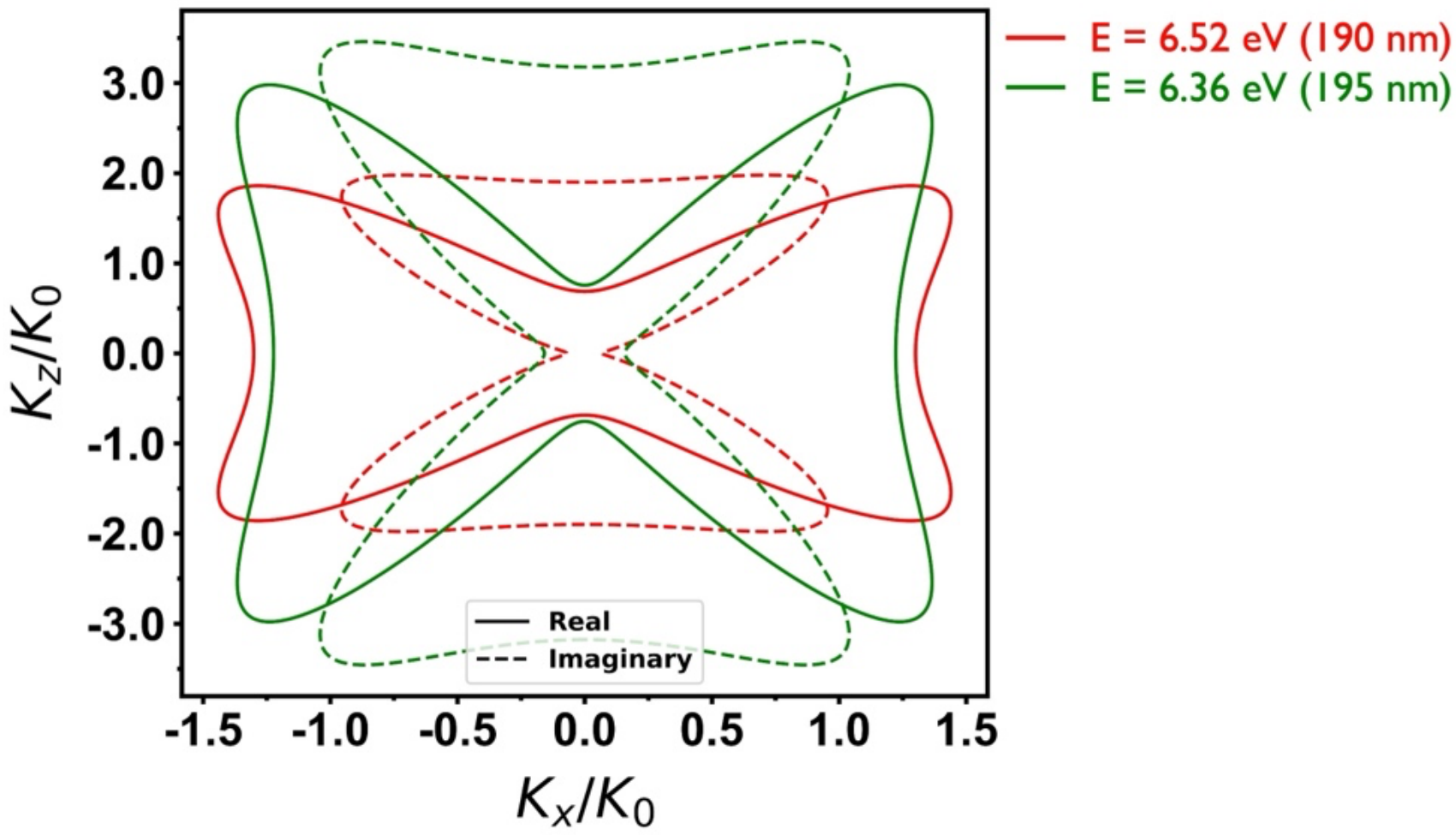

**Figure S9**. Isofrequency surface in various photon energies (wavelengths) of hBN in the hyperbolic regime, showing the hyperbolic dispersion.

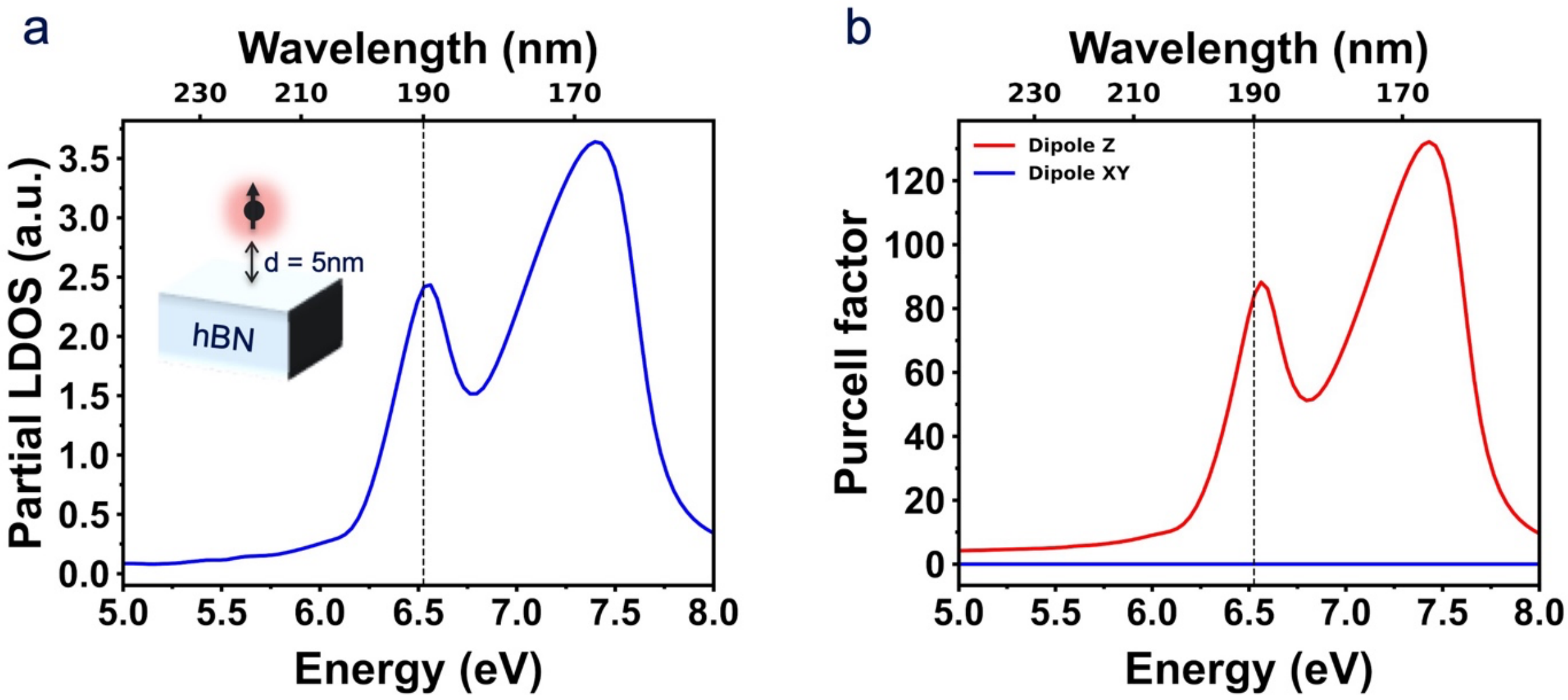

**Figure S10.** (a) Calculated partial local density of states (LDOS) using finite-difference time-domain (FDTD) simulations. The dipole along the z-axis is located 5 nm above the hBN, as illustrated in the inset. The partial LDOS can be calculated from the imaginary part of the Green's function, demonstrating the larger partial LDOS in the hyperbolic regime[17]. (b) Simulated Purcell factor depending on the dipole orientation under the same conditions in (a). When the dipole is oriented along the z direction, the significant Purcell factor can be achieved, consistent with the partial LDOS calculation, as the Purcell factor is proportional to the LDOS. In contrast, an in-plane-oriented dipole does not contribute to emission enhancement, highlighting the critical role of hyperbolic dispersion in boosting the Purcell factor in hBN. In addition, a notable Purcell factor can be achieved around the epsilon near-zero (ENZ) point[18,19].

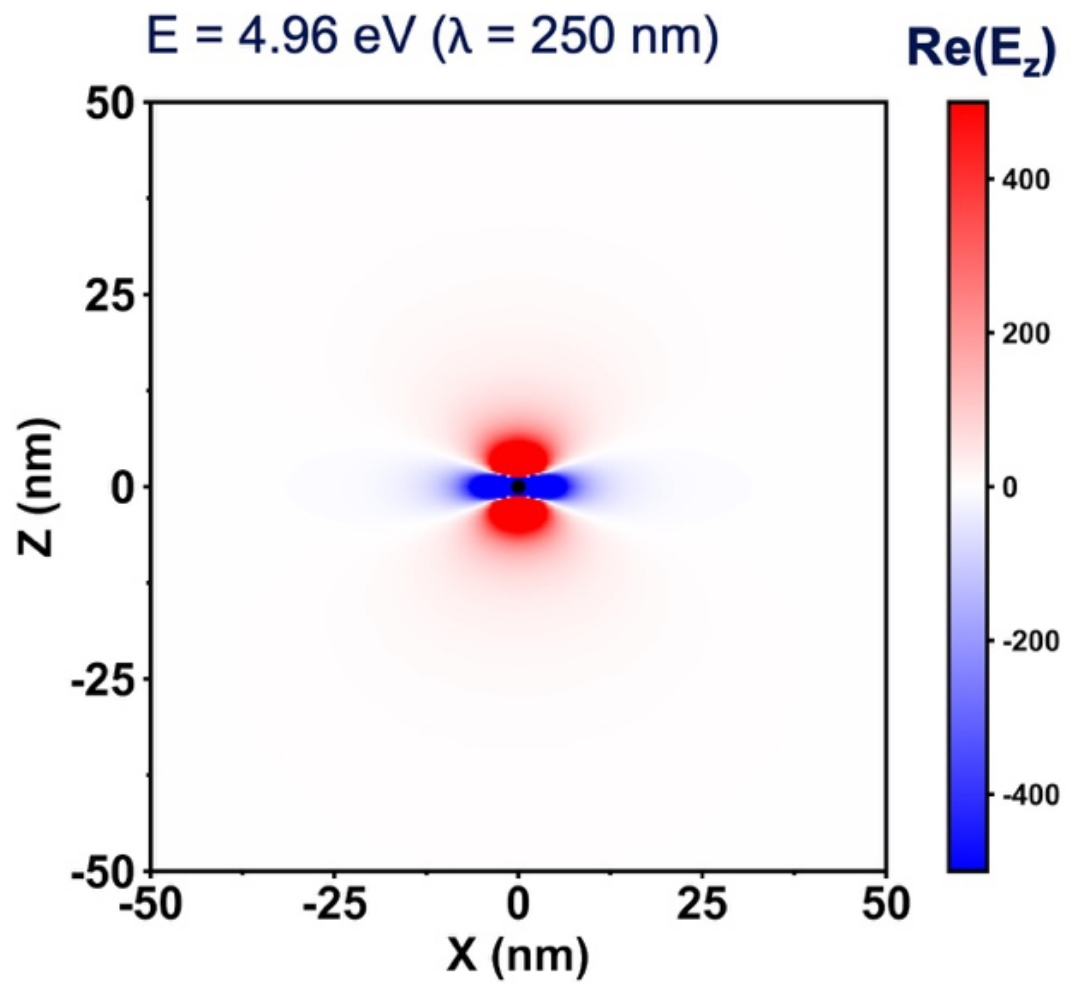

**Figure S11.** Numerically simulated near-field distribution obtained from FDTD simulations shows the wave launched by a point dipole oriented along the Z-axis, outside the hyperbolic regime, in the XZ plane, exhibiting non-directional propagation behavior.

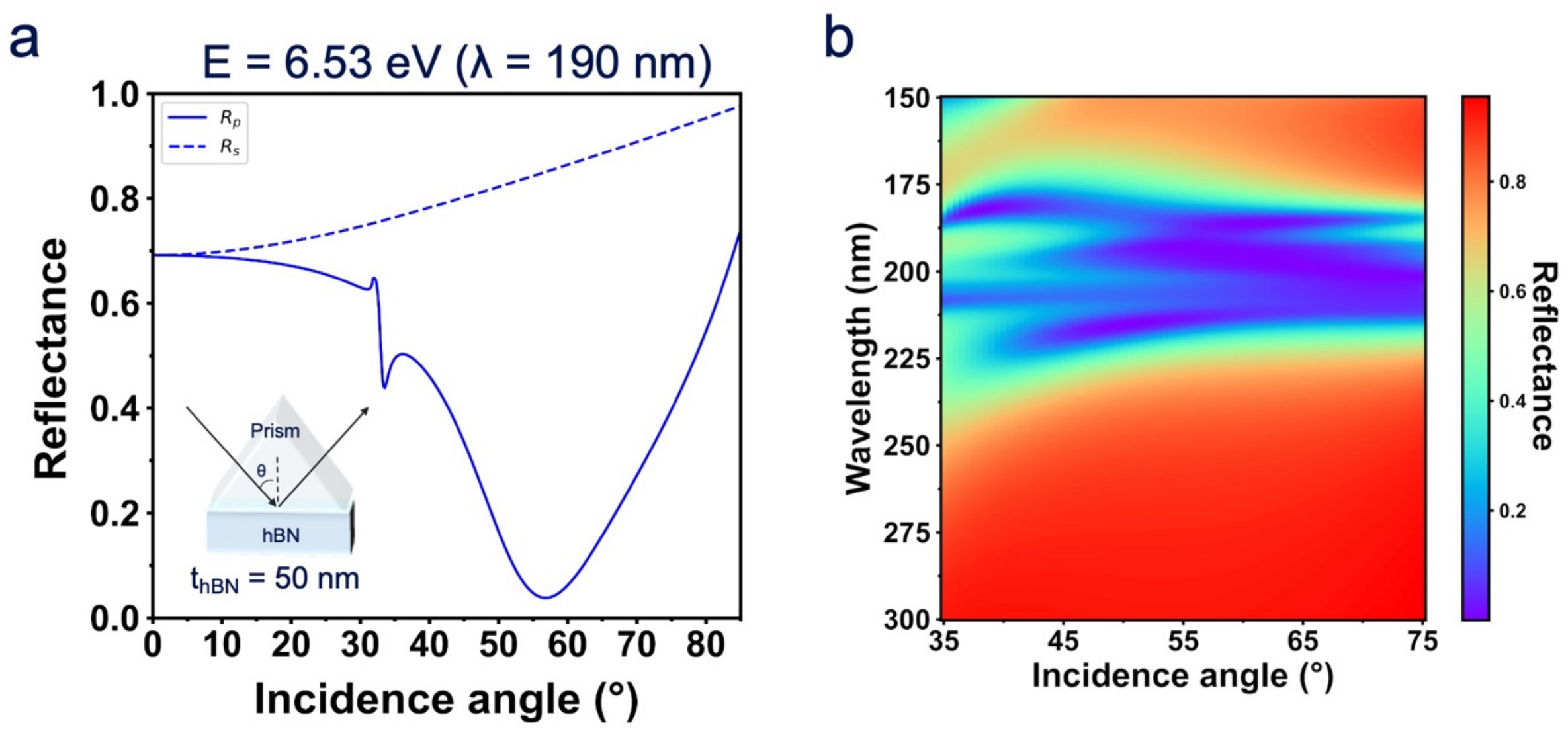

**Figure S12.** (a) The calculated reflectance spectrum of 50 nm of hBN as a function of incidence angle at E = 6.53 eV (λ = 190 nm), depending on the polarization of incidence light (s (TE) and p (TM) polarization). Otto configuration is adapted using a high-index prism ($Al_2O_3$) to launch the highly confined high-k mode[20]. TE polarized light shows high reflectance due to the existence of Reststrahlen in hBN. In contrast, the pronounced dip appears around a 30° incidence angle in the TM polarized light incidence due to the coupling with the high-k mode[20]. Furthermore, Brewster's angle is observed around a 60° incidence angle, consistent with theoretical predictions calculated from the dielectric constants of the media. (b) The simulated reflectance spectrum as a function of incidence angle and wavelength, indicating a clear high-k wave launch in the hyperbolic window.

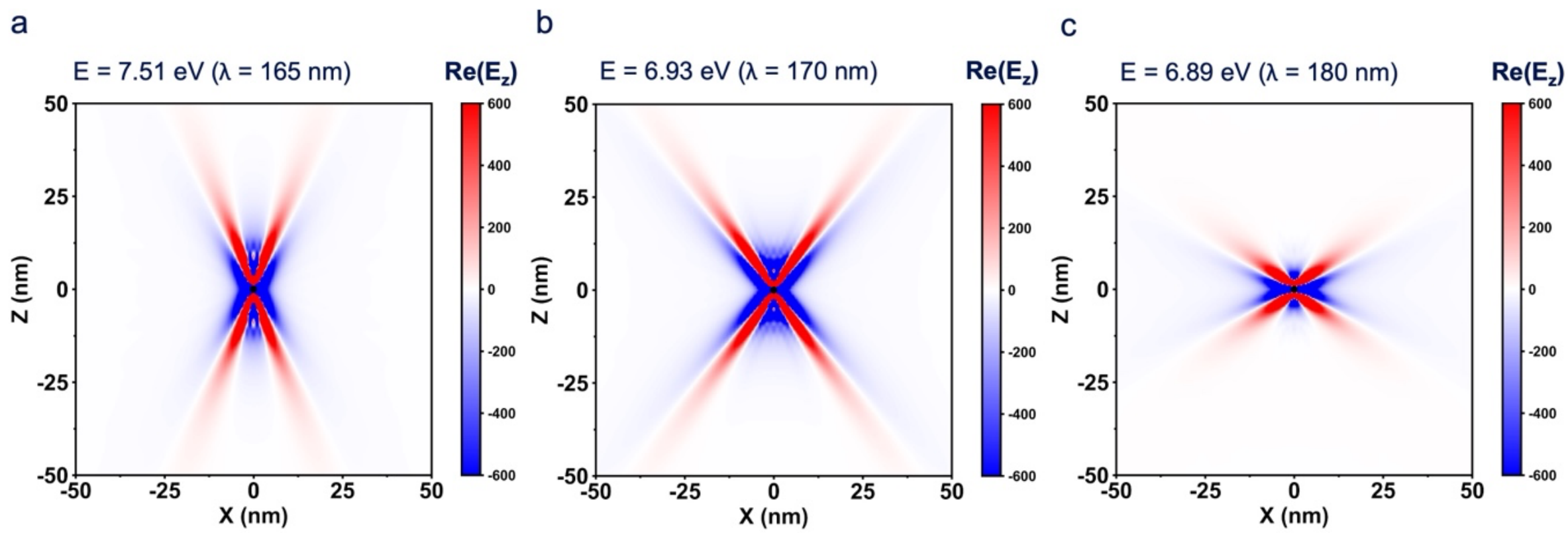

**Figure S13.** Numerically simulated near-field wave propagation obtained from FDTD simulations as a function of photon energies (wavelengths) (a) 7.51 eV (λ = 165 nm), (b) 6.93 eV (λ = 170 nm), and (c) 6.89 eV (λ = 180 nm), showing highly directional propagation depending on the photon energies (wavelengths) in hyperbolic regime.

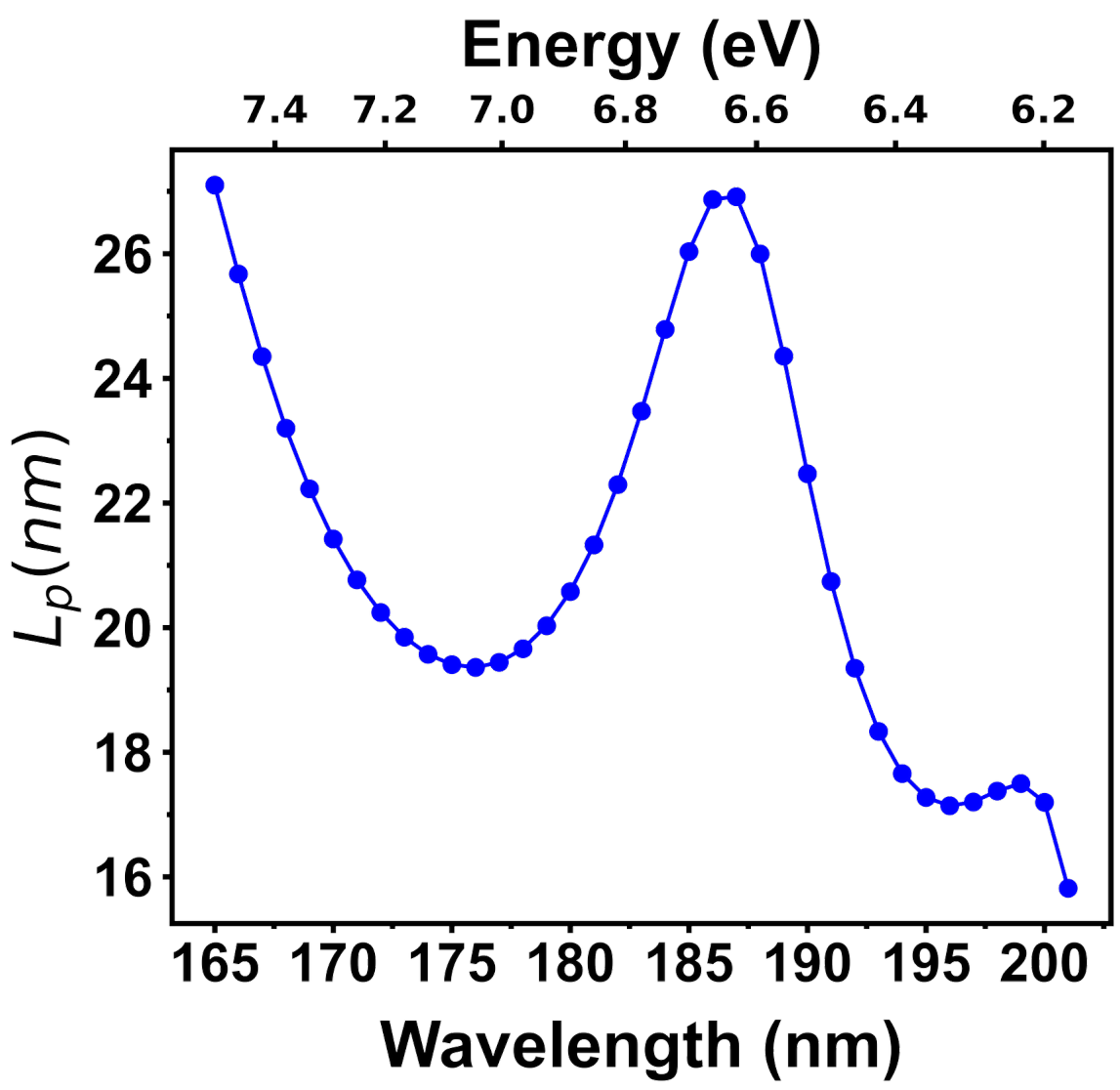

**Figure S14.** The propagation length of hyperbolic exciton polaritons (HEP) along the direction corresponding to the maximum real part of the wavevector in the hyperbolic regime[21].

**Table 1. Anisotropic, complex refractive index of hBN**

| wavelength (nm) | $n_{in\text{-}plane}$ ($n_o$) | $k_{in\text{-}plane}$ ($k_o$) | $n_{out\text{-}of\text{-}plane}$ ($n_e$) | $k_{out\text{-}of\text{-}plane}$ ($k_e$) |
|---|---|---|---|---|
| 190 | 0.6866 | 1.8992 | 1.3006 | 0.0704 |
| 191 | 0.6398 | 2.1088 | 1.2873 | 0.0814 |
| 192 | 0.6234 | 2.3385 | 1.2728 | 0.0948 |
| 193 | 0.6355 | 2.5894 | 1.2571 | 0.1114 |
| 194 | 0.6774 | 2.8663 | 1.2403 | 0.1321 |
| 195 | 0.7552 | 3.1769 | 1.2229 | 0.1582 |
| 196 | 0.882 | 3.5321 | 1.2057 | 0.1912 |
| 197 | 1.0838 | 3.9453 | 1.1904 | 0.233 |
| 198 | 1.4114 | 4.429 | 1.1801 | 0.2854 |
| 199 | 1.9646 | 4.9751 | 1.1797 | 0.3494 |
| 200 | 2.9206 | 5.4731 | 1.196 | 0.4237 |
| 201 | 4.3979 | 5.4884 | 1.2363 | 0.5027 |
| 202 | 5.7732 | 4.4482 | 1.3058 | 0.5758 |
| 203 | 6.1308 | 2.9839 | 1.4032 | 0.6285 |
| 204 | 5.8395 | 1.9483 | 1.5183 | 0.6467 |
| 205 | 5.4229 | 1.3385 | 1.6326 | 0.6244 |
| 206 | 5.0454 | 0.9747 | 1.7269 | 0.5681 |
| 207 | 4.7314 | 0.745 | 1.7908 | 0.4935 |
| 208 | 4.4735 | 0.5916 | 1.8255 | 0.4165 |
| 209 | 4.2602 | 0.484 | 1.838 | 0.3469 |
| 210 | 4.0815 | 0.4055 | 1.8366 | 0.2882 |
| 211 | 3.9298 | 0.3463 | 1.8273 | 0.2405 |
| 212 | 3.7996 | 0.3005 | 1.8141 | 0.2022 |
| 213 | 3.6865 | 0.2642 | 1.7994 | 0.1716 |
| 214 | 3.5873 | 0.2349 | 1.7846 | 0.1471 |
| 215 | 3.4996 | 0.2108 | 1.7702 | 0.1272 |
| 216 | 3.4214 | 0.1907 | 1.7567 | 0.111 |
| 217 | 3.3512 | 0.1738 | 1.7441 | 0.0976 |
| 218 | 3.2879 | 0.1593 | 1.7325 | 0.0866 |
| 219 | 3.2303 | 0.1469 | 1.7219 | 0.0773 |
| 220 | 3.1779 | 0.1361 | 1.7121 | 0.0694 |
| 221 | 3.1298 | 0.1266 | 1.7031 | 0.0627 |
| 222 | 3.0855 | 0.1183 | 1.6949 | 0.0569 |
| 223 | 3.0447 | 0.1109 | 1.6873 | 0.052 |
| 224 | 3.0068 | 0.1043 | 1.6803 | 0.0476 |
| 225 | 2.9716 | 0.0984 | 1.6738 | 0.0438 |
| 226 | 2.9388 | 0.0931 | 1.6678 | 0.0405 |
| 227 | 2.9081 | 0.0883 | 1.6623 | 0.0376 |
| 228 | 2.8794 | 0.0839 | 1.6571 | 0.0349 |
| 229 | 2.8524 | 0.08 | 1.6523 | 0.0326 |
| 230 | 2.827 | 0.0763 | 1.6478 | 0.0305 |
| 231 | 2.8031 | 0.073 | 1.6436 | 0.0286 |
| 232 | 2.7805 | 0.0699 | 1.6397 | 0.0269 |
| 233 | 2.7591 | 0.067 | 1.636 | 0.0253 |
| 234 | 2.7389 | 0.0644 | 1.6326 | 0.0239 |
| 235 | 2.7196 | 0.062 | 1.6293 | 0.0226 |
| 236 | 2.7013 | 0.0597 | 1.6263 | 0.0215 |
| 237 | 2.6839 | 0.0576 | 1.6234 | 0.0204 |
| 238 | 2.6674 | 0.0556 | 1.6206 | 0.0194 |
| 239 | 2.6515 | 0.0537 | 1.618 | 0.0185 |
| 240 | 2.6364 | 0.052 | 1.6156 | 0.0176 |
| 241 | 2.622 | 0.0504 | 1.6133 | 0.0168 |
| 242 | 2.6081 | 0.0488 | 1.611 | 0.0161 |
| 243 | 2.5949 | 0.0474 | 1.6089 | 0.0154 |
| 244 | 2.5822 | 0.046 | 1.6069 | 0.0148 |
| 245 | 2.5699 | 0.0447 | 1.605 | 0.0142 |
| 246 | 2.5582 | 0.0435 | 1.6032 | 0.0137 |
| 247 | 2.5469 | 0.0423 | 1.6014 | 0.0131 |
| 248 | 2.5361 | 0.0412 | 1.5997 | 0.0127 |
| 249 | 2.5256 | 0.0402 | 1.5981 | 0.0122 |
| 250 | 2.5155 | 0.0392 | 1.5966 | 0.0118 |
| 251 | 2.5058 | 0.0383 | 1.5951 | 0.0114 |
| 252 | 2.4964 | 0.0374 | 1.5937 | 0.011 |
| 253 | 2.4874 | 0.0365 | 1.5924 | 0.0106 |
| 254 | 2.4786 | 0.0357 | 1.5911 | 0.0103 |
| 255 | 2.4702 | 0.0349 | 1.5898 | 0.01 |
| 256 | 2.462 | 0.0341 | 1.5886 | 0.0097 |
| 257 | 2.454 | 0.0334 | 1.5875 | 0.0094 |
| 258 | 2.4464 | 0.0327 | 1.5863 | 0.0091 |
| 259 | 2.4389 | 0.0321 | 1.5853 | 0.0088 |
| 260 | 2.4317 | 0.0314 | 1.5842 | 0.0086 |
| 261 | 2.4247 | 0.0308 | 1.5832 | 0.0084 |
| 262 | 2.4179 | 0.0302 | 1.5822 | 0.0081 |
| 263 | 2.4113 | 0.0297 | 1.5813 | 0.0079 |
| 264 | 2.4049 | 0.0291 | 1.5804 | 0.0077 |
| 265 | 2.3987 | 0.0286 | 1.5795 | 0.0075 |
| 266 | 2.3926 | 0.0281 | 1.5787 | 0.0073 |
| 267 | 2.3867 | 0.0276 | 1.5778 | 0.0072 |
| 268 | 2.381 | 0.0272 | 1.577 | 0.007 |
| 269 | 2.3754 | 0.0267 | 1.5763 | 0.0068 |
| 270 | 2.37 | 0.0263 | 1.5755 | 0.0067 |
| 271 | 2.3647 | 0.0259 | 1.5748 | 0.0065 |
| 272 | 2.3596 | 0.0254 | 1.5741 | 0.0064 |
| 273 | 2.3546 | 0.025 | 1.5734 | 0.0062 |
| 274 | 2.3497 | 0.0247 | 1.5727 | 0.0061 |
| 275 | 2.3449 | 0.0243 | 1.5721 | 0.006 |
| 276 | 2.3403 | 0.0239 | 1.5714 | 0.0058 |
| 277 | 2.3357 | 0.0236 | 1.5708 | 0.0057 |
| 278 | 2.3313 | 0.0233 | 1.5702 | 0.0056 |
| 279 | 2.3269 | 0.0229 | 1.5696 | 0.0055 |
| 280 | 2.3227 | 0.0226 | 1.5691 | 0.0054 |
| 281 | 2.3186 | 0.0223 | 1.5685 | 0.0053 |
| 282 | 2.3146 | 0.022 | 1.568 | 0.0052 |
| 283 | 2.3106 | 0.0217 | 1.5675 | 0.0051 |
| 284 | 2.3068 | 0.0214 | 1.567 | 0.005 |
| 285 | 2.303 | 0.0211 | 1.5665 | 0.0049 |
| 286 | 2.2993 | 0.0209 | 1.566 | 0.0048 |
| 287 | 2.2957 | 0.0206 | 1.5655 | 0.0047 |
| 288 | 2.2921 | 0.0204 | 1.565 | 0.0047 |
| 289 | 2.2887 | 0.0201 | 1.5646 | 0.0046 |
| 290 | 2.2853 | 0.0199 | 1.5641 | 0.0045 |
| 291 | 2.282 | 0.0196 | 1.5637 | 0.0044 |
| 292 | 2.2787 | 0.0194 | 1.5633 | 0.0043 |
| 293 | 2.2756 | 0.0192 | 1.5629 | 0.0043 |
| 294 | 2.2724 | 0.019 | 1.5625 | 0.0042 |
| 295 | 2.2694 | 0.0188 | 1.5621 | 0.0041 |
| 296 | 2.2664 | 0.0186 | 1.5617 | 0.0041 |
| 297 | 2.2635 | 0.0183 | 1.5613 | 0.004 |
| 298 | 2.2606 | 0.0181 | 1.561 | 0.004 |
| 299 | 2.2578 | 0.018 | 1.5606 | 0.0039 |
| 300 | 2.255 | 0.0178 | 1.5602 | 0.0038 |
| 301 | 2.2523 | 0.0176 | 1.5599 | 0.0038 |
| 302 | 2.2496 | 0.0174 | 1.5596 | 0.0037 |
| 303 | 2.247 | 0.0172 | 1.5592 | 0.0037 |
| 304 | 2.2444 | 0.0171 | 1.5589 | 0.0036 |
| 305 | 2.2419 | 0.0169 | 1.5586 | 0.0036 |
| 306 | 2.2394 | 0.0167 | 1.5583 | 0.0035 |
| 307 | 2.237 | 0.0166 | 1.558 | 0.0035 |
| 308 | 2.2346 | 0.0164 | 1.5577 | 0.0034 |
| 309 | 2.2322 | 0.0162 | 1.5574 | 0.0034 |
| 310 | 2.2299 | 0.0161 | 1.5571 | 0.0033 |
| 311 | 2.2277 | 0.0159 | 1.5568 | 0.0033 |
| 312 | 2.2254 | 0.0158 | 1.5565 | 0.0033 |
| 313 | 2.2232 | 0.0157 | 1.5563 | 0.0032 |
| 314 | 2.2211 | 0.0155 | 1.556 | 0.0032 |
| 315 | 2.219 | 0.0154 | 1.5557 | 0.0031 |
| 316 | 2.2169 | 0.0152 | 1.5555 | 0.0031 |
| 317 | 2.2148 | 0.0151 | 1.5552 | 0.0031 |
| 318 | 2.2128 | 0.015 | 1.555 | 0.003 |
| 319 | 2.2108 | 0.0148 | 1.5547 | 0.003 |
| 320 | 2.2089 | 0.0147 | 1.5545 | 0.003 |
| 321 | 2.207 | 0.0146 | 1.5543 | 0.0029 |
| 322 | 2.2051 | 0.0145 | 1.554 | 0.0029 |
| 323 | 2.2032 | 0.0144 | 1.5538 | 0.0029 |
| 324 | 2.2014 | 0.0142 | 1.5536 | 0.0028 |
| 325 | 2.1996 | 0.0141 | 1.5534 | 0.0028 |
| 326 | 2.1978 | 0.014 | 1.5532 | 0.0028 |
| 327 | 2.1961 | 0.0139 | 1.5529 | 0.0027 |
| 328 | 2.1944 | 0.0138 | 1.5527 | 0.0027 |
| 329 | 2.1927 | 0.0137 | 1.5525 | 0.0027 |
| 330 | 2.191 | 0.0136 | 1.5523 | 0.0026 |
| 331 | 2.1894 | 0.0135 | 1.5521 | 0.0026 |
| 332 | 2.1877 | 0.0134 | 1.5519 | 0.0026 |
| 333 | 2.1861 | 0.0133 | 1.5517 | 0.0026 |
| 334 | 2.1846 | 0.0132 | 1.5516 | 0.0025 |
| 335 | 2.183 | 0.0131 | 1.5514 | 0.0025 |
| 336 | 2.1815 | 0.013 | 1.5512 | 0.0025 |
| 337 | 2.18 | 0.0129 | 1.551 | 0.0025 |
| 338 | 2.1785 | 0.0128 | 1.5508 | 0.0024 |
| 339 | 2.1771 | 0.0127 | 1.5507 | 0.0024 |
| 340 | 2.1756 | 0.0126 | 1.5505 | 0.0024 |
| 341 | 2.1742 | 0.0125 | 1.5503 | 0.0024 |
| 342 | 2.1728 | 0.0125 | 1.5502 | 0.0024 |
| 343 | 2.1714 | 0.0124 | 1.55 | 0.0023 |
| 344 | 2.17 | 0.0123 | 1.5498 | 0.0023 |
| 345 | 2.1687 | 0.0122 | 1.5497 | 0.0023 |
| 346 | 2.1674 | 0.0121 | 1.5495 | 0.0023 |
| 347 | 2.1661 | 0.012 | 1.5494 | 0.0022 |
| 348 | 2.1648 | 0.012 | 1.5492 | 0.0022 |
| 349 | 2.1635 | 0.0119 | 1.5491 | 0.0022 |
| 350 | 2.1622 | 0.0118 | 1.5489 | 0.0022 |
| 351 | 2.161 | 0.0117 | 1.5488 | 0.0022 |
| 352 | 2.1598 | 0.0117 | 1.5486 | 0.0022 |
| 353 | 2.1586 | 0.0116 | 1.5485 | 0.0021 |
| 354 | 2.1574 | 0.0115 | 1.5483 | 0.0021 |
| 355 | 2.1562 | 0.0115 | 1.5482 | 0.0021 |
| 356 | 2.155 | 0.0114 | 1.5481 | 0.0021 |
| 357 | 2.1539 | 0.0113 | 1.5479 | 0.0021 |
| 358 | 2.1528 | 0.0112 | 1.5478 | 0.002 |
| 359 | 2.1516 | 0.0112 | 1.5477 | 0.002 |
| 360 | 2.1505 | 0.0111 | 1.5475 | 0.002 |
| 361 | 2.1495 | 0.011 | 1.5474 | 0.002 |
| 362 | 2.1484 | 0.011 | 1.5473 | 0.002 |
| 363 | 2.1473 | 0.0109 | 1.5472 | 0.002 |
| 364 | 2.1463 | 0.0109 | 1.547 | 0.002 |
| 365 | 2.1452 | 0.0108 | 1.5469 | 0.0019 |
| 366 | 2.1442 | 0.0107 | 1.5468 | 0.0019 |
| 367 | 2.1432 | 0.0107 | 1.5467 | 0.0019 |
| 368 | 2.1422 | 0.0106 | 1.5466 | 0.0019 |
| 369 | 2.1412 | 0.0105 | 1.5465 | 0.0019 |
| 370 | 2.1402 | 0.0105 | 1.5463 | 0.0019 |
| 371 | 2.1393 | 0.0104 | 1.5462 | 0.0019 |
| 372 | 2.1383 | 0.0104 | 1.5461 | 0.0018 |
| 373 | 2.1374 | 0.0103 | 1.546 | 0.0018 |
| 374 | 2.1364 | 0.0103 | 1.5459 | 0.0018 |
| 375 | 2.1355 | 0.0102 | 1.5458 | 0.0018 |
| 376 | 2.1346 | 0.0102 | 1.5457 | 0.0018 |
| 377 | 2.1337 | 0.0101 | 1.5456 | 0.0018 |
| 378 | 2.1328 | 0.01 | 1.5455 | 0.0018 |
| 379 | 2.1319 | 0.01 | 1.5454 | 0.0017 |
| 380 | 2.131 | 0.0099 | 1.5453 | 0.0017 |
| 381 | 2.1302 | 0.0099 | 1.5452 | 0.0017 |
| 382 | 2.1293 | 0.0098 | 1.5451 | 0.0017 |
| 383 | 2.1285 | 0.0098 | 1.545 | 0.0017 |
| 384 | 2.1277 | 0.0097 | 1.5449 | 0.0017 |
| 385 | 2.1268 | 0.0097 | 1.5448 | 0.0017 |
| 386 | 2.126 | 0.0096 | 1.5447 | 0.0017 |
| 387 | 2.1252 | 0.0096 | 1.5446 | 0.0017 |
| 388 | 2.1244 | 0.0095 | 1.5445 | 0.0016 |
| 389 | 2.1236 | 0.0095 | 1.5444 | 0.0016 |
| 390 | 2.1229 | 0.0095 | 1.5444 | 0.0016 |
| 391 | 2.1221 | 0.0094 | 1.5443 | 0.0016 |
| 392 | 2.1213 | 0.0094 | 1.5442 | 0.0016 |
| 393 | 2.1206 | 0.0093 | 1.5441 | 0.0016 |
| 394 | 2.1198 | 0.0093 | 1.544 | 0.0016 |
| 395 | 2.1191 | 0.0092 | 1.5439 | 0.0016 |
| 396 | 2.1184 | 0.0092 | 1.5439 | 0.0016 |
| 397 | 2.1176 | 0.0091 | 1.5438 | 0.0016 |
| 398 | 2.1169 | 0.0091 | 1.5437 | 0.0015 |
| 399 | 2.1162 | 0.0091 | 1.5436 | 0.0015 |
| 400 | 2.1155 | 0.009 | 1.5435 | 0.0015 |
| 401 | 2.1148 | 0.009 | 1.5435 | 0.0015 |
| 402 | 2.1141 | 0.0089 | 1.5434 | 0.0015 |
| 403 | 2.1134 | 0.0089 | 1.5433 | 0.0015 |
| 404 | 2.1128 | 0.0089 | 1.5432 | 0.0015 |
| 405 | 2.1121 | 0.0088 | 1.5431 | 0.0015 |
| 406 | 2.1114 | 0.0088 | 1.5431 | 0.0015 |
| 407 | 2.1108 | 0.0087 | 1.543 | 0.0015 |
| 408 | 2.1101 | 0.0087 | 1.5429 | 0.0015 |
| 409 | 2.1095 | 0.0087 | 1.5429 | 0.0014 |
| 410 | 2.1089 | 0.0086 | 1.5428 | 0.0014 |
| 411 | 2.1082 | 0.0086 | 1.5427 | 0.0014 |
| 412 | 2.1076 | 0.0086 | 1.5426 | 0.0014 |
| 413 | 2.107 | 0.0085 | 1.5426 | 0.0014 |
| 414 | 2.1064 | 0.0085 | 1.5425 | 0.0014 |
| 415 | 2.1058 | 0.0084 | 1.5424 | 0.0014 |
| 416 | 2.1052 | 0.0084 | 1.5424 | 0.0014 |
| 417 | 2.1046 | 0.0084 | 1.5423 | 0.0014 |
| 418 | 2.104 | 0.0083 | 1.5422 | 0.0014 |
| 419 | 2.1034 | 0.0083 | 1.5422 | 0.0014 |
| 420 | 2.1029 | 0.0083 | 1.5421 | 0.0014 |
| 421 | 2.1023 | 0.0082 | 1.5421 | 0.0014 |
| 422 | 2.1017 | 0.0082 | 1.542 | 0.0014 |
| 423 | 2.1012 | 0.0082 | 1.5419 | 0.0013 |
| 424 | 2.1006 | 0.0081 | 1.5419 | 0.0013 |
| 425 | 2.1001 | 0.0081 | 1.5418 | 0.0013 |
| 426 | 2.0995 | 0.0081 | 1.5417 | 0.0013 |
| 427 | 2.099 | 0.008 | 1.5417 | 0.0013 |
| 428 | 2.0985 | 0.008 | 1.5416 | 0.0013 |
| 429 | 2.0979 | 0.008 | 1.5416 | 0.0013 |
| 430 | 2.0974 | 0.0079 | 1.5415 | 0.0013 |
| 431 | 2.0969 | 0.0079 | 1.5415 | 0.0013 |
| 432 | 2.0964 | 0.0079 | 1.5414 | 0.0013 |
| 433 | 2.0959 | 0.0079 | 1.5413 | 0.0013 |
| 434 | 2.0954 | 0.0078 | 1.5413 | 0.0013 |
| 435 | 2.0949 | 0.0078 | 1.5412 | 0.0013 |
| 436 | 2.0944 | 0.0078 | 1.5412 | 0.0013 |
| 437 | 2.0939 | 0.0077 | 1.5411 | 0.0013 |
| 438 | 2.0934 | 0.0077 | 1.5411 | 0.0012 |
| 439 | 2.0929 | 0.0077 | 1.541 | 0.0012 |
| 440 | 2.0924 | 0.0076 | 1.541 | 0.0012 |
| 441 | 2.0919 | 0.0076 | 1.5409 | 0.0012 |
| 442 | 2.0915 | 0.0076 | 1.5409 | 0.0012 |
| 443 | 2.091 | 0.0076 | 1.5408 | 0.0012 |
| 444 | 2.0905 | 0.0075 | 1.5408 | 0.0012 |
| 445 | 2.0901 | 0.0075 | 1.5407 | 0.0012 |
| 446 | 2.0896 | 0.0075 | 1.5407 | 0.0012 |
| 447 | 2.0892 | 0.0075 | 1.5406 | 0.0012 |
| 448 | 2.0887 | 0.0074 | 1.5406 | 0.0012 |
| 449 | 2.0883 | 0.0074 | 1.5405 | 0.0012 |
| 450 | 2.0878 | 0.0074 | 1.5405 | 0.0012 |
| 451 | 2.0874 | 0.0073 | 1.5404 | 0.0012 |
| 452 | 2.087 | 0.0073 | 1.5404 | 0.0012 |
| 453 | 2.0866 | 0.0073 | 1.5403 | 0.0012 |
| 454 | 2.0861 | 0.0073 | 1.5403 | 0.0012 |
| 455 | 2.0857 | 0.0072 | 1.5402 | 0.0012 |
| 456 | 2.0853 | 0.0072 | 1.5402 | 0.0011 |
| 457 | 2.0849 | 0.0072 | 1.5401 | 0.0011 |
| 458 | 2.0845 | 0.0072 | 1.5401 | 0.0011 |
| 459 | 2.0841 | 0.0071 | 1.54 | 0.0011 |
| 460 | 2.0837 | 0.0071 | 1.54 | 0.0011 |
| 461 | 2.0833 | 0.0071 | 1.54 | 0.0011 |
| 462 | 2.0829 | 0.0071 | 1.5399 | 0.0011 |
| 463 | 2.0825 | 0.007 | 1.5399 | 0.0011 |
| 464 | 2.0821 | 0.007 | 1.5398 | 0.0011 |
| 465 | 2.0817 | 0.007 | 1.5398 | 0.0011 |
| 466 | 2.0813 | 0.007 | 1.5397 | 0.0011 |
| 467 | 2.0809 | 0.007 | 1.5397 | 0.0011 |
| 468 | 2.0805 | 0.0069 | 1.5397 | 0.0011 |
| 469 | 2.0802 | 0.0069 | 1.5396 | 0.0011 |
| 470 | 2.0798 | 0.0069 | 1.5396 | 0.0011 |
| 471 | 2.0794 | 0.0069 | 1.5395 | 0.0011 |
| 472 | 2.0791 | 0.0068 | 1.5395 | 0.0011 |
| 473 | 2.0787 | 0.0068 | 1.5395 | 0.0011 |
| 474 | 2.0783 | 0.0068 | 1.5394 | 0.0011 |
| 475 | 2.078 | 0.0068 | 1.5394 | 0.0011 |
| 476 | 2.0776 | 0.0068 | 1.5393 | 0.0011 |
| 477 | 2.0773 | 0.0067 | 1.5393 | 0.0011 |
| 478 | 2.0769 | 0.0067 | 1.5393 | 0.001 |
| 479 | 2.0766 | 0.0067 | 1.5392 | 0.001 |
| 480 | 2.0762 | 0.0067 | 1.5392 | 0.001 |
| 481 | 2.0759 | 0.0066 | 1.5392 | 0.001 |
| 482 | 2.0755 | 0.0066 | 1.5391 | 0.001 |
| 483 | 2.0752 | 0.0066 | 1.5391 | 0.001 |
| 484 | 2.0749 | 0.0066 | 1.5391 | 0.001 |
| 485 | 2.0745 | 0.0066 | 1.539 | 0.001 |
| 486 | 2.0742 | 0.0065 | 1.539 | 0.001 |
| 487 | 2.0739 | 0.0065 | 1.5389 | 0.001 |
| 488 | 2.0736 | 0.0065 | 1.5389 | 0.001 |
| 489 | 2.0732 | 0.0065 | 1.5389 | 0.001 |
| 490 | 2.0729 | 0.0065 | 1.5388 | 0.001 |
| 491 | 2.0726 | 0.0064 | 1.5388 | 0.001 |
| 492 | 2.0723 | 0.0064 | 1.5388 | 0.001 |
| 493 | 2.072 | 0.0064 | 1.5387 | 0.001 |
| 494 | 2.0717 | 0.0064 | 1.5387 | 0.001 |
| 495 | 2.0714 | 0.0064 | 1.5387 | 0.001 |
| 496 | 2.0711 | 0.0063 | 1.5386 | 0.001 |
| 497 | 2.0708 | 0.0063 | 1.5386 | 0.001 |
| 498 | 2.0705 | 0.0063 | 1.5386 | 0.001 |
| 499 | 2.0702 | 0.0063 | 1.5385 | 0.001 |
| 500 | 2.0699 | 0.0063 | 1.5385 | 0.001 |
| 501 | 2.0696 | 0.0063 | 1.5385 | 0.001 |
| 502 | 2.0693 | 0.0062 | 1.5384 | 0.001 |
| 503 | 2.069 | 0.0062 | 1.5384 | 0.001 |
| 504 | 2.0687 | 0.0062 | 1.5384 | 0.001 |
| 505 | 2.0684 | 0.0062 | 1.5384 | 0.0009 |
| 506 | 2.0681 | 0.0062 | 1.5383 | 0.0009 |
| 507 | 2.0678 | 0.0061 | 1.5383 | 0.0009 |
| 508 | 2.0676 | 0.0061 | 1.5383 | 0.0009 |
| 509 | 2.0673 | 0.0061 | 1.5382 | 0.0009 |
| 510 | 2.067 | 0.0061 | 1.5382 | 0.0009 |
| 511 | 2.0667 | 0.0061 | 1.5382 | 0.0009 |
| 512 | 2.0665 | 0.0061 | 1.5381 | 0.0009 |
| 513 | 2.0662 | 0.006 | 1.5381 | 0.0009 |
| 514 | 2.0659 | 0.006 | 1.5381 | 0.0009 |
| 515 | 2.0657 | 0.006 | 1.5381 | 0.0009 |
| 516 | 2.0654 | 0.006 | 1.538 | 0.0009 |
| 517 | 2.0651 | 0.006 | 1.538 | 0.0009 |
| 518 | 2.0649 | 0.006 | 1.538 | 0.0009 |
| 519 | 2.0646 | 0.0059 | 1.5379 | 0.0009 |
| 520 | 2.0644 | 0.0059 | 1.5379 | 0.0009 |
| 521 | 2.0641 | 0.0059 | 1.5379 | 0.0009 |
| 522 | 2.0638 | 0.0059 | 1.5379 | 0.0009 |
| 523 | 2.0636 | 0.0059 | 1.5378 | 0.0009 |
| 524 | 2.0633 | 0.0059 | 1.5378 | 0.0009 |
| 525 | 2.0631 | 0.0059 | 1.5378 | 0.0009 |
| 526 | 2.0629 | 0.0058 | 1.5378 | 0.0009 |
| 527 | 2.0626 | 0.0058 | 1.5377 | 0.0009 |
| 528 | 2.0624 | 0.0058 | 1.5377 | 0.0009 |
| 529 | 2.0621 | 0.0058 | 1.5377 | 0.0009 |
| 530 | 2.0619 | 0.0058 | 1.5377 | 0.0009 |
| 531 | 2.0616 | 0.0058 | 1.5376 | 0.0009 |
| 532 | 2.0614 | 0.0057 | 1.5376 | 0.0009 |
| 533 | 2.0612 | 0.0057 | 1.5376 | 0.0009 |
| 534 | 2.0609 | 0.0057 | 1.5376 | 0.0009 |
| 535 | 2.0607 | 0.0057 | 1.5375 | 0.0009 |
| 536 | 2.0605 | 0.0057 | 1.5375 | 0.0009 |
| 537 | 2.0602 | 0.0057 | 1.5375 | 0.0009 |
| 538 | 2.06 | 0.0057 | 1.5375 | 0.0008 |
| 539 | 2.0598 | 0.0056 | 1.5374 | 0.0008 |
| 540 | 2.0596 | 0.0056 | 1.5374 | 0.0008 |
| 541 | 2.0593 | 0.0056 | 1.5374 | 0.0008 |
| 542 | 2.0591 | 0.0056 | 1.5374 | 0.0008 |
| 543 | 2.0589 | 0.0056 | 1.5373 | 0.0008 |
| 544 | 2.0587 | 0.0056 | 1.5373 | 0.0008 |
| 545 | 2.0585 | 0.0056 | 1.5373 | 0.0008 |
| 546 | 2.0582 | 0.0055 | 1.5373 | 0.0008 |
| 547 | 2.058 | 0.0055 | 1.5372 | 0.0008 |
| 548 | 2.0578 | 0.0055 | 1.5372 | 0.0008 |
| 549 | 2.0576 | 0.0055 | 1.5372 | 0.0008 |
| 550 | 2.0574 | 0.0055 | 1.5372 | 0.0008 |
| 551 | 2.0572 | 0.0055 | 1.5372 | 0.0008 |
| 552 | 2.057 | 0.0055 | 1.5371 | 0.0008 |
| 553 | 2.0568 | 0.0054 | 1.5371 | 0.0008 |
| 554 | 2.0566 | 0.0054 | 1.5371 | 0.0008 |
| 555 | 2.0564 | 0.0054 | 1.5371 | 0.0008 |
| 556 | 2.0562 | 0.0054 | 1.537 | 0.0008 |
| 557 | 2.056 | 0.0054 | 1.537 | 0.0008 |
| 558 | 2.0558 | 0.0054 | 1.537 | 0.0008 |
| 559 | 2.0556 | 0.0054 | 1.537 | 0.0008 |
| 560 | 2.0554 | 0.0054 | 1.537 | 0.0008 |
| 561 | 2.0552 | 0.0053 | 1.5369 | 0.0008 |
| 562 | 2.055 | 0.0053 | 1.5369 | 0.0008 |
| 563 | 2.0548 | 0.0053 | 1.5369 | 0.0008 |
| 564 | 2.0546 | 0.0053 | 1.5369 | 0.0008 |
| 565 | 2.0544 | 0.0053 | 1.5369 | 0.0008 |
| 566 | 2.0542 | 0.0053 | 1.5368 | 0.0008 |
| 567 | 2.054 | 0.0053 | 1.5368 | 0.0008 |
| 568 | 2.0538 | 0.0053 | 1.5368 | 0.0008 |
| 569 | 2.0536 | 0.0052 | 1.5368 | 0.0008 |
| 570 | 2.0534 | 0.0052 | 1.5368 | 0.0008 |
| 571 | 2.0533 | 0.0052 | 1.5367 | 0.0008 |
| 572 | 2.0531 | 0.0052 | 1.5367 | 0.0008 |
| 573 | 2.0529 | 0.0052 | 1.5367 | 0.0008 |
| 574 | 2.0527 | 0.0052 | 1.5367 | 0.0008 |
| 575 | 2.0525 | 0.0052 | 1.5367 | 0.0008 |
| 576 | 2.0524 | 0.0052 | 1.5366 | 0.0008 |
| 577 | 2.0522 | 0.0051 | 1.5366 | 0.0008 |
| 578 | 2.052 | 0.0051 | 1.5366 | 0.0008 |
| 579 | 2.0518 | 0.0051 | 1.5366 | 0.0008 |
| 580 | 2.0516 | 0.0051 | 1.5366 | 0.0008 |
| 581 | 2.0515 | 0.0051 | 1.5366 | 0.0008 |
| 582 | 2.0513 | 0.0051 | 1.5365 | 0.0007 |
| 583 | 2.0511 | 0.0051 | 1.5365 | 0.0007 |
| 584 | 2.051 | 0.0051 | 1.5365 | 0.0007 |
| 585 | 2.0508 | 0.005 | 1.5365 | 0.0007 |
| 586 | 2.0506 | 0.005 | 1.5365 | 0.0007 |
| 587 | 2.0505 | 0.005 | 1.5364 | 0.0007 |
| 588 | 2.0503 | 0.005 | 1.5364 | 0.0007 |
| 589 | 2.0501 | 0.005 | 1.5364 | 0.0007 |
| 590 | 2.05 | 0.005 | 1.5364 | 0.0007 |
| 591 | 2.0498 | 0.005 | 1.5364 | 0.0007 |
| 592 | 2.0496 | 0.005 | 1.5364 | 0.0007 |
| 593 | 2.0495 | 0.005 | 1.5363 | 0.0007 |
| 594 | 2.0493 | 0.0049 | 1.5363 | 0.0007 |
| 595 | 2.0491 | 0.0049 | 1.5363 | 0.0007 |
| 596 | 2.049 | 0.0049 | 1.5363 | 0.0007 |
| 597 | 2.0488 | 0.0049 | 1.5363 | 0.0007 |
| 598 | 2.0487 | 0.0049 | 1.5363 | 0.0007 |
| 599 | 2.0485 | 0.0049 | 1.5362 | 0.0007 |
| 600 | 2.0484 | 0.0049 | 1.5362 | 0.0007 |
| 601 | 2.0482 | 0.0049 | 1.5362 | 0.0007 |
| 602 | 2.048 | 0.0049 | 1.5362 | 0.0007 |
| 603 | 2.0479 | 0.0049 | 1.5362 | 0.0007 |
| 604 | 2.0477 | 0.0048 | 1.5362 | 0.0007 |
| 605 | 2.0476 | 0.0048 | 1.5361 | 0.0007 |
| 606 | 2.0474 | 0.0048 | 1.5361 | 0.0007 |
| 607 | 2.0473 | 0.0048 | 1.5361 | 0.0007 |
| 608 | 2.0471 | 0.0048 | 1.5361 | 0.0007 |
| 609 | 2.047 | 0.0048 | 1.5361 | 0.0007 |
| 610 | 2.0468 | 0.0048 | 1.5361 | 0.0007 |
| 611 | 2.0467 | 0.0048 | 1.536 | 0.0007 |
| 612 | 2.0466 | 0.0048 | 1.536 | 0.0007 |
| 613 | 2.0464 | 0.0048 | 1.536 | 0.0007 |
| 614 | 2.0463 | 0.0047 | 1.536 | 0.0007 |
| 615 | 2.0461 | 0.0047 | 1.536 | 0.0007 |
| 616 | 2.046 | 0.0047 | 1.536 | 0.0007 |
| 617 | 2.0458 | 0.0047 | 1.536 | 0.0007 |
| 618 | 2.0457 | 0.0047 | 1.5359 | 0.0007 |
| 619 | 2.0456 | 0.0047 | 1.5359 | 0.0007 |
| 620 | 2.0454 | 0.0047 | 1.5359 | 0.0007 |
| 621 | 2.0453 | 0.0047 | 1.5359 | 0.0007 |
| 622 | 2.0451 | 0.0047 | 1.5359 | 0.0007 |
| 623 | 2.045 | 0.0047 | 1.5359 | 0.0007 |
| 624 | 2.0449 | 0.0046 | 1.5359 | 0.0007 |
| 625 | 2.0447 | 0.0046 | 1.5358 | 0.0007 |
| 626 | 2.0446 | 0.0046 | 1.5358 | 0.0007 |
| 627 | 2.0445 | 0.0046 | 1.5358 | 0.0007 |
| 628 | 2.0443 | 0.0046 | 1.5358 | 0.0007 |
| 629 | 2.0442 | 0.0046 | 1.5358 | 0.0007 |
| 630 | 2.0441 | 0.0046 | 1.5358 | 0.0007 |
| 631 | 2.0439 | 0.0046 | 1.5358 | 0.0007 |
| 632 | 2.0438 | 0.0046 | 1.5357 | 0.0007 |
| 633 | 2.0437 | 0.0046 | 1.5357 | 0.0007 |
| 634 | 2.0435 | 0.0046 | 1.5357 | 0.0007 |
| 635 | 2.0434 | 0.0045 | 1.5357 | 0.0007 |
| 636 | 2.0433 | 0.0045 | 1.5357 | 0.0007 |
| 637 | 2.0432 | 0.0045 | 1.5357 | 0.0007 |
| 638 | 2.043 | 0.0045 | 1.5357 | 0.0007 |
| 639 | 2.0429 | 0.0045 | 1.5357 | 0.0007 |
| 640 | 2.0428 | 0.0045 | 1.5356 | 0.0006 |
| 641 | 2.0427 | 0.0045 | 1.5356 | 0.0006 |
| 642 | 2.0425 | 0.0045 | 1.5356 | 0.0006 |
| 643 | 2.0424 | 0.0045 | 1.5356 | 0.0006 |
| 644 | 2.0423 | 0.0045 | 1.5356 | 0.0006 |
| 645 | 2.0422 | 0.0045 | 1.5356 | 0.0006 |
| 646 | 2.042 | 0.0044 | 1.5356 | 0.0006 |
| 647 | 2.0419 | 0.0044 | 1.5355 | 0.0006 |
| 648 | 2.0418 | 0.0044 | 1.5355 | 0.0006 |
| 649 | 2.0417 | 0.0044 | 1.5355 | 0.0006 |
| 650 | 2.0416 | 0.0044 | 1.5355 | 0.0006 |
| 651 | 2.0414 | 0.0044 | 1.5355 | 0.0006 |
| 652 | 2.0413 | 0.0044 | 1.5355 | 0.0006 |
| 653 | 2.0412 | 0.0044 | 1.5355 | 0.0006 |
| 654 | 2.0411 | 0.0044 | 1.5355 | 0.0006 |
| 655 | 2.041 | 0.0044 | 1.5355 | 0.0006 |
| 656 | 2.0409 | 0.0044 | 1.5354 | 0.0006 |
| 657 | 2.0407 | 0.0044 | 1.5354 | 0.0006 |
| 658 | 2.0406 | 0.0043 | 1.5354 | 0.0006 |
| 659 | 2.0405 | 0.0043 | 1.5354 | 0.0006 |
| 660 | 2.0404 | 0.0043 | 1.5354 | 0.0006 |
| 661 | 2.0403 | 0.0043 | 1.5354 | 0.0006 |
| 662 | 2.0402 | 0.0043 | 1.5354 | 0.0006 |
| 663 | 2.0401 | 0.0043 | 1.5354 | 0.0006 |
| 664 | 2.04 | 0.0043 | 1.5353 | 0.0006 |
| 665 | 2.0398 | 0.0043 | 1.5353 | 0.0006 |
| 666 | 2.0397 | 0.0043 | 1.5353 | 0.0006 |
| 667 | 2.0396 | 0.0043 | 1.5353 | 0.0006 |
| 668 | 2.0395 | 0.0043 | 1.5353 | 0.0006 |
| 669 | 2.0394 | 0.0043 | 1.5353 | 0.0006 |
| 670 | 2.0393 | 0.0042 | 1.5353 | 0.0006 |
| 671 | 2.0392 | 0.0042 | 1.5353 | 0.0006 |
| 672 | 2.0391 | 0.0042 | 1.5353 | 0.0006 |
| 673 | 2.039 | 0.0042 | 1.5352 | 0.0006 |
| 674 | 2.0389 | 0.0042 | 1.5352 | 0.0006 |
| 675 | 2.0388 | 0.0042 | 1.5352 | 0.0006 |
| 676 | 2.0387 | 0.0042 | 1.5352 | 0.0006 |
| 677 | 2.0386 | 0.0042 | 1.5352 | 0.0006 |
| 678 | 2.0385 | 0.0042 | 1.5352 | 0.0006 |
| 679 | 2.0384 | 0.0042 | 1.5352 | 0.0006 |
| 680 | 2.0383 | 0.0042 | 1.5352 | 0.0006 |
| 681 | 2.0382 | 0.0042 | 1.5352 | 0.0006 |
| 682 | 2.0381 | 0.0042 | 1.5351 | 0.0006 |
| 683 | 2.038 | 0.0042 | 1.5351 | 0.0006 |
| 684 | 2.0379 | 0.0041 | 1.5351 | 0.0006 |
| 685 | 2.0378 | 0.0041 | 1.5351 | 0.0006 |
| 686 | 2.0377 | 0.0041 | 1.5351 | 0.0006 |
| 687 | 2.0376 | 0.0041 | 1.5351 | 0.0006 |
| 688 | 2.0375 | 0.0041 | 1.5351 | 0.0006 |
| 689 | 2.0374 | 0.0041 | 1.5351 | 0.0006 |
| 690 | 2.0373 | 0.0041 | 1.5351 | 0.0006 |
| 691 | 2.0372 | 0.0041 | 1.5351 | 0.0006 |
| 692 | 2.0371 | 0.0041 | 1.535 | 0.0006 |
| 693 | 2.037 | 0.0041 | 1.535 | 0.0006 |
| 694 | 2.0369 | 0.0041 | 1.535 | 0.0006 |
| 695 | 2.0368 | 0.0041 | 1.535 | 0.0006 |
| 696 | 2.0367 | 0.0041 | 1.535 | 0.0006 |
| 697 | 2.0366 | 0.004 | 1.535 | 0.0006 |
| 698 | 2.0365 | 0.004 | 1.535 | 0.0006 |
| 699 | 2.0364 | 0.004 | 1.535 | 0.0006 |
| 700 | 2.0363 | 0.004 | 1.535 | 0.0006 |
| 701 | 2.0362 | 0.004 | 1.535 | 0.0006 |
| 702 | 2.0361 | 0.004 | 1.5349 | 0.0006 |
| 703 | 2.036 | 0.004 | 1.5349 | 0.0006 |
| 704 | 2.0359 | 0.004 | 1.5349 | 0.0006 |
| 705 | 2.0359 | 0.004 | 1.5349 | 0.0006 |
| 706 | 2.0358 | 0.004 | 1.5349 | 0.0006 |
| 707 | 2.0357 | 0.004 | 1.5349 | 0.0006 |
| 708 | 2.0356 | 0.004 | 1.5349 | 0.0006 |
| 709 | 2.0355 | 0.004 | 1.5349 | 0.0006 |
| 710 | 2.0354 | 0.004 | 1.5349 | 0.0006 |
| 711 | 2.0353 | 0.004 | 1.5349 | 0.0006 |
| 712 | 2.0352 | 0.0039 | 1.5349 | 0.0006 |
| 713 | 2.0351 | 0.0039 | 1.5348 | 0.0006 |
| 714 | 2.0351 | 0.0039 | 1.5348 | 0.0006 |
| 715 | 2.035 | 0.0039 | 1.5348 | 0.0006 |
| 716 | 2.0349 | 0.0039 | 1.5348 | 0.0006 |
| 717 | 2.0348 | 0.0039 | 1.5348 | 0.0006 |
| 718 | 2.0347 | 0.0039 | 1.5348 | 0.0006 |
| 719 | 2.0346 | 0.0039 | 1.5348 | 0.0006 |
| 720 | 2.0345 | 0.0039 | 1.5348 | 0.0006 |
| 721 | 2.0345 | 0.0039 | 1.5348 | 0.0005 |
| 722 | 2.0344 | 0.0039 | 1.5348 | 0.0005 |
| 723 | 2.0343 | 0.0039 | 1.5348 | 0.0005 |
| 724 | 2.0342 | 0.0039 | 1.5347 | 0.0005 |
| 725 | 2.0341 | 0.0039 | 1.5347 | 0.0005 |
| 726 | 2.034 | 0.0039 | 1.5347 | 0.0005 |
| 727 | 2.034 | 0.0038 | 1.5347 | 0.0005 |
| 728 | 2.0339 | 0.0038 | 1.5347 | 0.0005 |
| 729 | 2.0338 | 0.0038 | 1.5347 | 0.0005 |
| 730 | 2.0337 | 0.0038 | 1.5347 | 0.0005 |
| 731 | 2.0336 | 0.0038 | 1.5347 | 0.0005 |
| 732 | 2.0336 | 0.0038 | 1.5347 | 0.0005 |
| 733 | 2.0335 | 0.0038 | 1.5347 | 0.0005 |
| 734 | 2.0334 | 0.0038 | 1.5347 | 0.0005 |
| 735 | 2.0333 | 0.0038 | 1.5347 | 0.0005 |
| 736 | 2.0332 | 0.0038 | 1.5346 | 0.0005 |
| 737 | 2.0332 | 0.0038 | 1.5346 | 0.0005 |
| 738 | 2.0331 | 0.0038 | 1.5346 | 0.0005 |
| 739 | 2.033 | 0.0038 | 1.5346 | 0.0005 |
| 740 | 2.0329 | 0.0038 | 1.5346 | 0.0005 |
| 741 | 2.0329 | 0.0038 | 1.5346 | 0.0005 |
| 742 | 2.0328 | 0.0038 | 1.5346 | 0.0005 |
| 743 | 2.0327 | 0.0038 | 1.5346 | 0.0005 |
| 744 | 2.0326 | 0.0037 | 1.5346 | 0.0005 |
| 745 | 2.0325 | 0.0037 | 1.5346 | 0.0005 |
| 746 | 2.0325 | 0.0037 | 1.5346 | 0.0005 |
| 747 | 2.0324 | 0.0037 | 1.5346 | 0.0005 |
| 748 | 2.0323 | 0.0037 | 1.5346 | 0.0005 |
| 749 | 2.0322 | 0.0037 | 1.5345 | 0.0005 |
| 750 | 2.0322 | 0.0037 | 1.5345 | 0.0005 |
| 751 | 2.0321 | 0.0037 | 1.5345 | 0.0005 |
| 752 | 2.032 | 0.0037 | 1.5345 | 0.0005 |
| 753 | 2.032 | 0.0037 | 1.5345 | 0.0005 |
| 754 | 2.0319 | 0.0037 | 1.5345 | 0.0005 |
| 755 | 2.0318 | 0.0037 | 1.5345 | 0.0005 |
| 756 | 2.0317 | 0.0037 | 1.5345 | 0.0005 |
| 757 | 2.0317 | 0.0037 | 1.5345 | 0.0005 |
| 758 | 2.0316 | 0.0037 | 1.5345 | 0.0005 |
| 759 | 2.0315 | 0.0037 | 1.5345 | 0.0005 |
| 760 | 2.0315 | 0.0037 | 1.5345 | 0.0005 |
| 761 | 2.0314 | 0.0036 | 1.5345 | 0.0005 |
| 762 | 2.0313 | 0.0036 | 1.5344 | 0.0005 |
| 763 | 2.0312 | 0.0036 | 1.5344 | 0.0005 |
| 764 | 2.0312 | 0.0036 | 1.5344 | 0.0005 |
| 765 | 2.0311 | 0.0036 | 1.5344 | 0.0005 |
| 766 | 2.031 | 0.0036 | 1.5344 | 0.0005 |
| 767 | 2.031 | 0.0036 | 1.5344 | 0.0005 |
| 768 | 2.0309 | 0.0036 | 1.5344 | 0.0005 |
| 769 | 2.0308 | 0.0036 | 1.5344 | 0.0005 |
| 770 | 2.0308 | 0.0036 | 1.5344 | 0.0005 |
| 771 | 2.0307 | 0.0036 | 1.5344 | 0.0005 |
| 772 | 2.0306 | 0.0036 | 1.5344 | 0.0005 |
| 773 | 2.0306 | 0.0036 | 1.5344 | 0.0005 |
| 774 | 2.0305 | 0.0036 | 1.5344 | 0.0005 |
| 775 | 2.0304 | 0.0036 | 1.5344 | 0.0005 |
| 776 | 2.0304 | 0.0036 | 1.5344 | 0.0005 |
| 777 | 2.0303 | 0.0036 | 1.5343 | 0.0005 |
| 778 | 2.0302 | 0.0036 | 1.5343 | 0.0005 |
| 779 | 2.0302 | 0.0035 | 1.5343 | 0.0005 |
| 780 | 2.0301 | 0.0035 | 1.5343 | 0.0005 |
| 781 | 2.03 | 0.0035 | 1.5343 | 0.0005 |
| 782 | 2.03 | 0.0035 | 1.5343 | 0.0005 |
| 783 | 2.0299 | 0.0035 | 1.5343 | 0.0005 |
| 784 | 2.0298 | 0.0035 | 1.5343 | 0.0005 |
| 785 | 2.0298 | 0.0035 | 1.5343 | 0.0005 |
| 786 | 2.0297 | 0.0035 | 1.5343 | 0.0005 |
| 787 | 2.0296 | 0.0035 | 1.5343 | 0.0005 |
| 788 | 2.0296 | 0.0035 | 1.5343 | 0.0005 |
| 789 | 2.0295 | 0.0035 | 1.5343 | 0.0005 |
| 790 | 2.0295 | 0.0035 | 1.5343 | 0.0005 |
| 791 | 2.0294 | 0.0035 | 1.5343 | 0.0005 |
| 792 | 2.0293 | 0.0035 | 1.5342 | 0.0005 |
| 793 | 2.0293 | 0.0035 | 1.5342 | 0.0005 |
| 794 | 2.0292 | 0.0035 | 1.5342 | 0.0005 |
| 795 | 2.0291 | 0.0035 | 1.5342 | 0.0005 |
| 796 | 2.0291 | 0.0035 | 1.5342 | 0.0005 |
| 797 | 2.029 | 0.0035 | 1.5342 | 0.0005 |
| 798 | 2.029 | 0.0035 | 1.5342 | 0.0005 |
| 799 | 2.0289 | 0.0034 | 1.5342 | 0.0005 |
| 800 | 2.0288 | 0.0034 | 1.5342 | 0.0005 |
| 801 | 2.0288 | 0.0034 | 1.5342 | 0.0005 |
| 802 | 2.0287 | 0.0034 | 1.5342 | 0.0005 |
| 803 | 2.0287 | 0.0034 | 1.5342 | 0.0005 |
| 804 | 2.0286 | 0.0034 | 1.5342 | 0.0005 |
| 805 | 2.0285 | 0.0034 | 1.5342 | 0.0005 |
| 806 | 2.0285 | 0.0034 | 1.5342 | 0.0005 |
| 807 | 2.0284 | 0.0034 | 1.5342 | 0.0005 |
| 808 | 2.0284 | 0.0034 | 1.5341 | 0.0005 |
| 809 | 2.0283 | 0.0034 | 1.5341 | 0.0005 |
| 810 | 2.0283 | 0.0034 | 1.5341 | 0.0005 |
| 811 | 2.0282 | 0.0034 | 1.5341 | 0.0005 |
| 812 | 2.0281 | 0.0034 | 1.5341 | 0.0005 |
| 813 | 2.0281 | 0.0034 | 1.5341 | 0.0005 |
| 814 | 2.028 | 0.0034 | 1.5341 | 0.0005 |
| 815 | 2.028 | 0.0034 | 1.5341 | 0.0005 |
| 816 | 2.0279 | 0.0034 | 1.5341 | 0.0005 |
| 817 | 2.0279 | 0.0034 | 1.5341 | 0.0005 |
| 818 | 2.0278 | 0.0034 | 1.5341 | 0.0005 |
| 819 | 2.0277 | 0.0033 | 1.5341 | 0.0005 |
| 820 | 2.0277 | 0.0033 | 1.5341 | 0.0005 |
| 821 | 2.0276 | 0.0033 | 1.5341 | 0.0005 |
| 822 | 2.0276 | 0.0033 | 1.5341 | 0.0005 |
| 823 | 2.0275 | 0.0033 | 1.5341 | 0.0005 |
| 824 | 2.0275 | 0.0033 | 1.5341 | 0.0005 |
| 825 | 2.0274 | 0.0033 | 1.534 | 0.0005 |
| 826 | 2.0274 | 0.0033 | 1.534 | 0.0005 |
| 827 | 2.0273 | 0.0033 | 1.534 | 0.0005 |
| 828 | 2.0273 | 0.0033 | 1.534 | 0.0005 |
| 829 | 2.0272 | 0.0033 | 1.534 | 0.0005 |
| 830 | 2.0271 | 0.0033 | 1.534 | 0.0005 |
| 831 | 2.0271 | 0.0033 | 1.534 | 0.0005 |
| 832 | 2.027 | 0.0033 | 1.534 | 0.0005 |
| 833 | 2.027 | 0.0033 | 1.534 | 0.0005 |
| 834 | 2.0269 | 0.0033 | 1.534 | 0.0005 |
| 835 | 2.0269 | 0.0033 | 1.534 | 0.0005 |
| 836 | 2.0268 | 0.0033 | 1.534 | 0.0005 |
| 837 | 2.0268 | 0.0033 | 1.534 | 0.0005 |
| 838 | 2.0267 | 0.0033 | 1.534 | 0.0005 |
| 839 | 2.0267 | 0.0033 | 1.534 | 0.0005 |
| 840 | 2.0266 | 0.0033 | 1.534 | 0.0005 |
| 841 | 2.0266 | 0.0032 | 1.534 | 0.0004 |
| 842 | 2.0265 | 0.0032 | 1.534 | 0.0004 |
| 843 | 2.0265 | 0.0032 | 1.534 | 0.0004 |
| 844 | 2.0264 | 0.0032 | 1.5339 | 0.0004 |
| 845 | 2.0264 | 0.0032 | 1.5339 | 0.0004 |
| 846 | 2.0263 | 0.0032 | 1.5339 | 0.0004 |
| 847 | 2.0263 | 0.0032 | 1.5339 | 0.0004 |
| 848 | 2.0262 | 0.0032 | 1.5339 | 0.0004 |
| 849 | 2.0262 | 0.0032 | 1.5339 | 0.0004 |
| 850 | 2.0261 | 0.0032 | 1.5339 | 0.0004 |
| 851 | 2.0261 | 0.0032 | 1.5339 | 0.0004 |
| 852 | 2.026 | 0.0032 | 1.5339 | 0.0004 |
| 853 | 2.026 | 0.0032 | 1.5339 | 0.0004 |
| 854 | 2.0259 | 0.0032 | 1.5339 | 0.0004 |
| 855 | 2.0259 | 0.0032 | 1.5339 | 0.0004 |
| 856 | 2.0258 | 0.0032 | 1.5339 | 0.0004 |
| 857 | 2.0258 | 0.0032 | 1.5339 | 0.0004 |
| 858 | 2.0257 | 0.0032 | 1.5339 | 0.0004 |
| 859 | 2.0257 | 0.0032 | 1.5339 | 0.0004 |
| 860 | 2.0256 | 0.0032 | 1.5339 | 0.0004 |
| 861 | 2.0256 | 0.0032 | 1.5339 | 0.0004 |
| 862 | 2.0255 | 0.0032 | 1.5339 | 0.0004 |
| 863 | 2.0255 | 0.0032 | 1.5339 | 0.0004 |
| 864 | 2.0254 | 0.0032 | 1.5338 | 0.0004 |
| 865 | 2.0254 | 0.0031 | 1.5338 | 0.0004 |
| 866 | 2.0254 | 0.0031 | 1.5338 | 0.0004 |
| 867 | 2.0253 | 0.0031 | 1.5338 | 0.0004 |
| 868 | 2.0253 | 0.0031 | 1.5338 | 0.0004 |
| 869 | 2.0252 | 0.0031 | 1.5338 | 0.0004 |
| 870 | 2.0252 | 0.0031 | 1.5338 | 0.0004 |
| 871 | 2.0251 | 0.0031 | 1.5338 | 0.0004 |
| 872 | 2.0251 | 0.0031 | 1.5338 | 0.0004 |
| 873 | 2.025 | 0.0031 | 1.5338 | 0.0004 |
| 874 | 2.025 | 0.0031 | 1.5338 | 0.0004 |
| 875 | 2.0249 | 0.0031 | 1.5338 | 0.0004 |
| 876 | 2.0249 | 0.0031 | 1.5338 | 0.0004 |
| 877 | 2.0249 | 0.0031 | 1.5338 | 0.0004 |
| 878 | 2.0248 | 0.0031 | 1.5338 | 0.0004 |
| 879 | 2.0248 | 0.0031 | 1.5338 | 0.0004 |
| 880 | 2.0247 | 0.0031 | 1.5338 | 0.0004 |
| 881 | 2.0247 | 0.0031 | 1.5338 | 0.0004 |
| 882 | 2.0246 | 0.0031 | 1.5338 | 0.0004 |
| 883 | 2.0246 | 0.0031 | 1.5338 | 0.0004 |
| 884 | 2.0245 | 0.0031 | 1.5338 | 0.0004 |
| 885 | 2.0245 | 0.0031 | 1.5337 | 0.0004 |
| 886 | 2.0245 | 0.0031 | 1.5337 | 0.0004 |
| 887 | 2.0244 | 0.0031 | 1.5337 | 0.0004 |
| 888 | 2.0244 | 0.0031 | 1.5337 | 0.0004 |
| 889 | 2.0243 | 0.0031 | 1.5337 | 0.0004 |
| 890 | 2.0243 | 0.003 | 1.5337 | 0.0004 |
| 891 | 2.0242 | 0.003 | 1.5337 | 0.0004 |
| 892 | 2.0242 | 0.003 | 1.5337 | 0.0004 |
| 893 | 2.0242 | 0.003 | 1.5337 | 0.0004 |
| 894 | 2.0241 | 0.003 | 1.5337 | 0.0004 |
| 895 | 2.0241 | 0.003 | 1.5337 | 0.0004 |
| 896 | 2.024 | 0.003 | 1.5337 | 0.0004 |
| 897 | 2.024 | 0.003 | 1.5337 | 0.0004 |
| 898 | 2.0239 | 0.003 | 1.5337 | 0.0004 |
| 899 | 2.0239 | 0.003 | 1.5337 | 0.0004 |
| 900 | 2.0239 | 0.003 | 1.5337 | 0.0004 |
| 901 | 2.0238 | 0.003 | 1.5337 | 0.0004 |
| 902 | 2.0238 | 0.003 | 1.5337 | 0.0004 |
| 903 | 2.0237 | 0.003 | 1.5337 | 0.0004 |
| 904 | 2.0237 | 0.003 | 1.5337 | 0.0004 |
| 905 | 2.0237 | 0.003 | 1.5337 | 0.0004 |
| 906 | 2.0236 | 0.003 | 1.5337 | 0.0004 |
| 907 | 2.0236 | 0.003 | 1.5337 | 0.0004 |
| 908 | 2.0235 | 0.003 | 1.5337 | 0.0004 |
| 909 | 2.0235 | 0.003 | 1.5336 | 0.0004 |
| 910 | 2.0235 | 0.003 | 1.5336 | 0.0004 |
| 911 | 2.0234 | 0.003 | 1.5336 | 0.0004 |
| 912 | 2.0234 | 0.003 | 1.5336 | 0.0004 |
| 913 | 2.0233 | 0.003 | 1.5336 | 0.0004 |
| 914 | 2.0233 | 0.003 | 1.5336 | 0.0004 |
| 915 | 2.0233 | 0.003 | 1.5336 | 0.0004 |
| 916 | 2.0232 | 0.0029 | 1.5336 | 0.0004 |
| 917 | 2.0232 | 0.0029 | 1.5336 | 0.0004 |
| 918 | 2.0231 | 0.0029 | 1.5336 | 0.0004 |
| 919 | 2.0231 | 0.0029 | 1.5336 | 0.0004 |
| 920 | 2.0231 | 0.0029 | 1.5336 | 0.0004 |
| 921 | 2.023 | 0.0029 | 1.5336 | 0.0004 |
| 922 | 2.023 | 0.0029 | 1.5336 | 0.0004 |
| 923 | 2.023 | 0.0029 | 1.5336 | 0.0004 |
| 924 | 2.0229 | 0.0029 | 1.5336 | 0.0004 |
| 925 | 2.0229 | 0.0029 | 1.5336 | 0.0004 |
| 926 | 2.0228 | 0.0029 | 1.5336 | 0.0004 |
| 927 | 2.0228 | 0.0029 | 1.5336 | 0.0004 |
| 928 | 2.0228 | 0.0029 | 1.5336 | 0.0004 |
| 929 | 2.0227 | 0.0029 | 1.5336 | 0.0004 |
| 930 | 2.0227 | 0.0029 | 1.5336 | 0.0004 |
| 931 | 2.0227 | 0.0029 | 1.5336 | 0.0004 |
| 932 | 2.0226 | 0.0029 | 1.5336 | 0.0004 |
| 933 | 2.0226 | 0.0029 | 1.5336 | 0.0004 |
| 934 | 2.0225 | 0.0029 | 1.5335 | 0.0004 |
| 935 | 2.0225 | 0.0029 | 1.5335 | 0.0004 |
| 936 | 2.0225 | 0.0029 | 1.5335 | 0.0004 |
| 937 | 2.0224 | 0.0029 | 1.5335 | 0.0004 |
| 938 | 2.0224 | 0.0029 | 1.5335 | 0.0004 |
| 939 | 2.0224 | 0.0029 | 1.5335 | 0.0004 |
| 940 | 2.0223 | 0.0029 | 1.5335 | 0.0004 |
| 941 | 2.0223 | 0.0029 | 1.5335 | 0.0004 |
| 942 | 2.0223 | 0.0029 | 1.5335 | 0.0004 |
| 943 | 2.0222 | 0.0029 | 1.5335 | 0.0004 |
| 944 | 2.0222 | 0.0029 | 1.5335 | 0.0004 |
| 945 | 2.0222 | 0.0028 | 1.5335 | 0.0004 |
| 946 | 2.0221 | 0.0028 | 1.5335 | 0.0004 |
| 947 | 2.0221 | 0.0028 | 1.5335 | 0.0004 |
| 948 | 2.022 | 0.0028 | 1.5335 | 0.0004 |
| 949 | 2.022 | 0.0028 | 1.5335 | 0.0004 |
| 950 | 2.022 | 0.0028 | 1.5335 | 0.0004 |
| 951 | 2.0219 | 0.0028 | 1.5335 | 0.0004 |
| 952 | 2.0219 | 0.0028 | 1.5335 | 0.0004 |
| 953 | 2.0219 | 0.0028 | 1.5335 | 0.0004 |
| 954 | 2.0218 | 0.0028 | 1.5335 | 0.0004 |
| 955 | 2.0218 | 0.0028 | 1.5335 | 0.0004 |
| 956 | 2.0218 | 0.0028 | 1.5335 | 0.0004 |
| 957 | 2.0217 | 0.0028 | 1.5335 | 0.0004 |
| 958 | 2.0217 | 0.0028 | 1.5335 | 0.0004 |
| 959 | 2.0217 | 0.0028 | 1.5335 | 0.0004 |
| 960 | 2.0216 | 0.0028 | 1.5335 | 0.0004 |
| 961 | 2.0216 | 0.0028 | 1.5335 | 0.0004 |
| 962 | 2.0216 | 0.0028 | 1.5334 | 0.0004 |
| 963 | 2.0215 | 0.0028 | 1.5334 | 0.0004 |
| 964 | 2.0215 | 0.0028 | 1.5334 | 0.0004 |
| 965 | 2.0215 | 0.0028 | 1.5334 | 0.0004 |
| 966 | 2.0214 | 0.0028 | 1.5334 | 0.0004 |
| 967 | 2.0214 | 0.0028 | 1.5334 | 0.0004 |
| 968 | 2.0214 | 0.0028 | 1.5334 | 0.0004 |
| 969 | 2.0213 | 0.0028 | 1.5334 | 0.0004 |
| 970 | 2.0213 | 0.0028 | 1.5334 | 0.0004 |
| 971 | 2.0213 | 0.0028 | 1.5334 | 0.0004 |
| 972 | 2.0212 | 0.0028 | 1.5334 | 0.0004 |
| 973 | 2.0212 | 0.0028 | 1.5334 | 0.0004 |
| 974 | 2.0212 | 0.0028 | 1.5334 | 0.0004 |
| 975 | 2.0211 | 0.0028 | 1.5334 | 0.0004 |
| 976 | 2.0211 | 0.0027 | 1.5334 | 0.0004 |
| 977 | 2.0211 | 0.0027 | 1.5334 | 0.0004 |
| 978 | 2.021 | 0.0027 | 1.5334 | 0.0004 |
| 979 | 2.021 | 0.0027 | 1.5334 | 0.0004 |
| 980 | 2.021 | 0.0027 | 1.5334 | 0.0004 |
| 981 | 2.021 | 0.0027 | 1.5334 | 0.0004 |
| 982 | 2.0209 | 0.0027 | 1.5334 | 0.0004 |
| 983 | 2.0209 | 0.0027 | 1.5334 | 0.0004 |
| 984 | 2.0209 | 0.0027 | 1.5334 | 0.0004 |
| 985 | 2.0208 | 0.0027 | 1.5334 | 0.0004 |
| 986 | 2.0208 | 0.0027 | 1.5334 | 0.0004 |
| 987 | 2.0208 | 0.0027 | 1.5334 | 0.0004 |
| 988 | 2.0207 | 0.0027 | 1.5334 | 0.0004 |
| 989 | 2.0207 | 0.0027 | 1.5334 | 0.0004 |
| 990 | 2.0207 | 0.0027 | 1.5334 | 0.0004 |
| 991 | 2.0206 | 0.0027 | 1.5334 | 0.0004 |
| 992 | 2.0206 | 0.0027 | 1.5334 | 0.0004 |
| 993 | 2.0206 | 0.0027 | 1.5333 | 0.0004 |
| 994 | 2.0206 | 0.0027 | 1.5333 | 0.0004 |
| 995 | 2.0205 | 0.0027 | 1.5333 | 0.0004 |
| 996 | 2.0205 | 0.0027 | 1.5333 | 0.0004 |
| 997 | 2.0205 | 0.0027 | 1.5333 | 0.0004 |
| 998 | 2.0204 | 0.0027 | 1.5333 | 0.0004 |
| 999 | 2.0204 | 0.0027 | 1.5333 | 0.0004 |
| 1000 | 2.0204 | 0.0027 | 1.5333 | 0.0004 |